\documentclass[apj]{aastex631}
\usepackage{graphicx}	
\usepackage{amsmath}	
\usepackage{amssymb}	
\usepackage{breqn}
\usepackage{natbib}
\usepackage{mathtools}
\renewcommand{\vec}[1]{\boldsymbol{#1}}
\newcommand{\pd}[2]{\frac{\partial #1}{\partial #2}}
\newcommand{\HALF}{\frac{1}{2}}
\newcommand{\DS}{\displaystyle}

\shorttitle{Numerical modeling and physical interplay of stochastic turbulent acceleration}
\shortauthors{Kundu, Vaidya, Mignone}
\graphicspath{{./}{figures/}}
\begin{document}
\title{Numerical modeling and physical interplay of stochastic turbulent acceleration for non-thermal emission processes.}
\author{Sayan Kundu}
\affiliation{Discipline of Astronomy, Astrophysics and Space Engineering \\
Indian Institute of Technology, Indore \\
Madhya Pradesh, India - 452020}

\author{Bhargav Vaidya}
\affiliation{Discipline of Astronomy, Astrophysics and Space Engineering \\
Indian Institute of Technology, Indore \\
Madhya Pradesh, India - 452020}

\author{Andrea Mignone}
\affiliation{Dipartimento di Fisica Generale, Universita degli Studi di Torino \\
Via Pietro Giuria 1, 10125 Torino, Italy}

\email{sayan.astronomy@gmail.com}
\begin{abstract}
Particle acceleration is an ubiquitous phenomenon in astrophysical and space plasma. 
Diffusive shock acceleration (DSA) and stochastic turbulent acceleration are known to be the possible mechanisms for producing very high energetic particles, particularly in weakly magnetized regions.
An interplay of different acceleration processes along with various radiation losses is typically observed in astrophysical sources. 
While DSA is a systematic acceleration process that energizes particles in the vicinity of shocks, stochastic turbulent acceleration (STA) is a random energizing process, where the interaction between cosmic ray particles and electromagnetic fluctuations results in particle acceleration.
This process is usually interpreted as a biased random walk in energy space, modelled through a Fokker-Planck equation. 
In the present work, we describe a novel Eulerian algorithm, adopted to 
incorporate turbulent acceleration in the presence of DSA and radiative processes like synchrotron and Inverse-Compton emission. 
The developed framework extends the hybrid Eulerian-Lagrangian module in a full-fledged relativistic Magneto-hydrodynamic (RMHD) code PLUTO. 
From our validation tests and case studies, we showcase the competing and complementary nature of both acceleration processes. Axisymmetric simulations of an RMHD jet with this extended hybrid framework clearly demonstrate that emission due to shocks is localized while that due to turbulent acceleration originates in the backflow and is more diffuse, particularly in the high energy X-ray band. 
\end{abstract}
%
\section{Introduction}
%
%

From giving a universal power-law trend to the cosmic ray spectrum to explaining the observed emission features of various astrophysical sources, particle acceleration process plays a crucial role in shaping our understanding of the nature of various space and astrophysical phenomena. 
Several observations require particles to be accelerated to very high energies in order to explain the energetics in different astrophysical sources. Due to high electrical conductivity, astrophysical plasma is incapable of sustaining a global electric field, making it challenging to energize particles in this scenario. 
Particle acceleration processes provide an alternative way to accelerate particles in the absence of a global electric field. The existing literature \citep{blandford_1994,kirk_1994,melrose_1996} suggests three main approaches to accelerate charged particles in an astrophysical plasma environment: shock acceleration (DSA), coherent electric field acceleration, and stochastic acceleration (STA). 

In \cite{fermi_1949}, Fermi first gave a proper mechanism for accelerating charged particles to explain the cosmic ray spectrum and the possible origin of high-energy cosmic ray particles. 
The mechanism considers relativistic particles getting scattered by moving inhomogeneities, mainly various plasma waves (MHD waves for highly relativistic cosmic ray particles \citep{parker_1955,Sturrock_1966,Kulsrud_1971}), and gaining energy (accelerate) in a randomized manner. 
This process is known as stochastic turbulent acceleration (STA) process. 
The randomness in the acceleration makes this process inefficient to energize particles, as suggested by the emission timescales observed in various astrophysical sources. 
Nevertheless, STA is considered to be an important source of turbulence damping in plasma and because of the omnipresence of turbulence in various astrophysical sources, STA has been invoked in order to explain the particle acceleration process in solar flares \citep{Petrosian_2012}, corona above accretion disk of compact object \citep{Dermer_1996,Liu_2004,Belmont_2008,Vurm_2009}, supernova remnant \citep{Bykov_1992,Kirk_1996,Macrowith_2010,Ferrand_2010},  gamma-ray burst \citep{Schlickeiser_2000}, emission from blazars(see \cite{Asano_2018} and references therein), radio lobes of AGN Jets \citep{sullivan_2009}, the diffuse X-ray emission from AGN jets \citep{fan_2008} along with fermi bubbles of galaxies \citep{Mertsch_2019}, galaxy clusters \citep{brunetti_2007,donnert_2014}. 
Recently STA has also been suggested as a candidate for the spectral gradient observed in galaxy clusters \citep{Rajpurohit_2020}. 

On the other hand, DSA gives a proper framework where particles can interact with the magnetic inhomogeneities in a way that could only increase the particles’ energy \citep{bell1_1978,drury_1983,blandford_1987, Malkov_2001}. 
Due to it's efficiency, DSA has been used to describe the particle acceleration process in various astrophysical systems, for example interplanetary helio-spheric shocks \citep{jokipii_2007,perri_2015}, shock wave of supernova remnant \citep{bell_2014}, stellar bow shock \citep{Rangelov_2019}, oblique shock in AGN jets \citep{meli_2013}, radio relics of galaxy clusters \citep{Kang_2017,van_weeren_2017,zimbardo_2017}. 
Though DSA is more efficient compared to STA mechanism, it is believed to only give rise to localized emission where STA is thought to produce large scale diffusive emission \citep{fan_2008}.  

To study these particle acceleration processes in various astrophysical systems, a numerical approach is imperative because of the multi-scale nature of the astrophysical plasma. 
Numerical study for plasma systems can broadly be categorized into different classes. 
Direct computation, mainly known as Particle in Cell (PIC) method, where Newton-Lorenz force law is solved along with Maxwell’s equation describing the dynamical evolution of the electric and magnetic field \citep{ Giacalone_2000, Nishikawa_2007, spitkovsky_2008,Sironi_2011}. 
This first principle approach has been taken by various researchers to study the particle acceleration processes \citep{comisso_2018,wong_2019, Marcowith_2020}. 
The next numerical scheme studies the plasma by solving the Vlasov equation for particle distribution evolution along with Maxwell’s equations \citep{Palmroth_2018}.
This scheme provides the advantage to study various plasma behaviour distinctively. 
This approach also enables us to study particle acceleration processes in different physical settings.
Similar to this approach, another approach is often taken to study particle acceleration process in the quasi-linear approximation where a Fokker-Plank equation is solved in order to evolve the cosmic ray spectrum due to interaction with MHD waves \citep{miniati_2001,donnert_2014,winner_2019,vazza_2021}.

Another numerical procedure studies the plasma in the fluid regime, also known as magneto-hydrodynamic (MHD) regime. This numerical procedure assumes plasma to be sufficiently collisional. 
That is why this procedure is incapable of capturing the physics of particle acceleration because collisions would make them to follow a Maxwellian which is in contrast to the observed power-law trend for the distribution of the accelerated particles. 
Though fluid approach fails to capture the particle acceleration process, it provides the background for the particles to interact with various MHD waves and accelerate. Recently some research has been devoted to combine the fluid and the PIC approaches \citep{Bai_2015} to study the DSA \citep{Mignone_2018}.
The final numerical method uses Monte-Carlo technique to study particle acceleration by shock wave \citep{achterberg_1992,baring_1994,marcowith_1999,wolff_2015} and turbulence \citep{Giacalone_1999,Teraki_2019}. 
Among all the numerical techniques available the Particle in Cell method has an advantage \citep{Ostrowski_1988,ellison_1990,ellison_2002,Lemoine_2003,Baring_2004,Niemiec_2006} over all other techniques because PIC not only can model the particle acceleration process, it also determine the self-generated magnetic turbulence, and treat them self-consistently with the cosmic ray particles. 
But the disadvantage of the PIC technique is, it is computationally very expensive \citep{Ellison_2013}.
And in order to bypass this problem other numerical techniques are used. 
Among them the kinetic test particle approach is one of the most efficient one because it could easily be incorporated with multi-scale simulations. 

As most of the sources of particle acceleration act simultaneously in different regions of astrophysical sources, it is imperative to develop a framework that can study such region to understand role of individual acceleration process. 
In this work, we use the kinetic test particle approach to study the competing and complimentary actions of DSA and STA. 
Other complimentary approaches have focused on studying the role either of the acceleration processes individually, for example,  \cite{Miniati_2001b,miniati_2003,donnert_2014} have demonstrated the role of STA in large scale galaxy clusters.

Recently, the existing Lagrangian particle module developed by \cite{vaidya_2018} in the PLUTO Code \citep{mignone_2007} has been applied to AGN jets at kpc scales to study the impact of instabilities and subsequent shocks on particle acceleration and non-thermal emission \citep{borse_2021, mukherjee_2021}. 
In the present work, we extend this Lagrangian framework by incorporating the STA process, to study the effect of both DSA and STA along with their roles in shaping the emission structure in astrophysical sources.
In this context, a macro-particle is a Lagrangian entity that moves along with the fluid and collects an ensemble of real particles (e.g. leptons) that are distributed in 1D momentum space.

The paper is organised as follows; in section \ref{sec:turb_theory}, we discuss the fundamental theory and necessary equations to describe the STA process. 
In section \ref{sec:IMEX}, we propose and describe a numerical algorithm to solve the cosmic ray transport equation.
We validate our algorithm and discuss it's accuracy in section \ref{sec:validation}. 
We analyze STA process in presence and absence of shocks in section \ref{sec:with_shock} and also discuss the role of several STA parameters through applications to test situations. 
Section \ref{sec:dis} discusses our findings and summarizes this work.       
%
\section{Turbulent Particle Acceleration : Theory} \label{sec:turb_theory}
%
%

This paper aims to study the effect of MHD turbulence and shocks on cosmic ray transport and their effect on the spectral signature of various astrophysical systems. The process of interaction between cosmic ray particles and turbulent plasma is stochastic in nature. Due to the random nature of the interaction, the energy of a cosmic ray particle follows a biased random walk, which leads the particle distribution to follow a diffusion equation \citep{tverskoi_1967}:

\begin{dmath}
\frac{{\partial} f_{0}}{{\partial} t} = \frac{1}{p^{2}} \frac{{\partial} }{{\partial} p} \left( p^{2} D_{pp} \frac{{\partial} f_{0}}{{\partial} p} \right)= \frac{\partial}{\partial p}\Bigg(D_{pp}\frac{\partial f_{0}}{\partial p}\Bigg) +\frac{2D_{pp}}{p}\frac{\partial f_0}{\partial p},
\label{eq:fermi_diff}
\end{dmath}

where, $f_{0}$ is the particle distribution function that depends on time $t$ and momentum $p$. $D_{pp}$ is the diffusion coefficient in momentum space. 
The above equation resembles a Fokker-Planck equation \citep{blandford_1987}. 
In a magnetized medium charged cosmic rays are also prone to loose their energy via various radiative and adiabatic losses. 
Inclusion of these loss effects along with the random interactions with turbulent magnetic fields results in the evolution of the distribution of relativistic cosmic ray particles as follows \citep{webb_1989}, 

\begin{dmath}{\label{eq:webb}}
    \nabla_{\mu}(u^{\mu} f_{0}+q^{\mu})+\frac{1}{p^{2}}\frac{\partial}{\partial p}\Big[-\frac{p^{3}}{3}f_{0}\nabla_{\mu} u^{\mu} + \langle \dot p\rangle_{L} f_{0} - \Gamma_{visc}p^{4}\tau \frac{\partial f_{0}}{\partial p} 
    - p^{2} D_{pp}\frac{\partial f_{0}}{\partial p} - p(p^{0})^{2} \dot u_{\mu} q^{\mu} \Big]=0.
\end{dmath}

The various terms of the equation are described below: 
\begin{enumerate}
    \item \label{term:1} $\nabla_{\mu}(u^{\mu} f_{0}+q^{\mu})$ represents the change in $f_{0}$, due to the spatial transport. $q^{\mu}$ is the spatial diffusion flux, $u^{\mu}$ is the bulk four-velocity;
    \item \label{term:2} $\frac{p^{3}}{3}f_{0}\nabla_{\mu} u^{\mu}$ defines the energy loss due to adiabatic expansion;
    \item \label{term:3} $\langle \dot p\rangle_{L} f_{0}$ describes the radiative losses, such as synchrotron and various Inverse Compton (IC) processes;
    \item \label{term:4} $\Gamma_{\rm visc}p^{4}\tau \frac{\partial f_{0}}{\partial p}$ is the particle acceleration term due to fluid shear \citep{Rieger_2019};
    \item \label{term:5} $p^{2} D_{pp}\frac{\partial f_{0}}{\partial p}$ represents the Fermi $\rm II$ order particle acceleration or STA process (see Eq.~(\ref{eq:fermi_diff}));
    \item \label{term:6} $p(p^{0})^{2} \dot u_{\mu} q^{\mu}$ originates because of the frame transformation.
\end{enumerate}

Following \cite{vaidya_2018}, we neglect the spatial diffusion flux $q^{\mu}$ as well as the acceleration due to frame transformation (i.e., terms ~\ref{term:1} and \ref{term:6}). 
Also, acceleration due to shear flow ($\Gamma_{\rm visc} =0)$ is not considered in the present study. 
Furthermore, the omission of the spatial diffusion term is compromised by an inclusion of a momentum independent escape term in Eq.~(\ref{eq:webb}) \citep{achterberg_1992}, so that Eq.~(\ref{eq:webb}) takes the form,

\begin{dmath}\label{eq:webb_1}
    \nabla_{\mu}(u^{\mu} f_{0})+\frac{1}{p^{2}}\frac{\partial}{\partial p}\Big[-\frac{p^{3}}{3}f_{0}\nabla_{\mu} u^{\mu} + \langle \dot p\rangle_{L} f_{0} - p^{2} D_{pp}\frac{\partial f_{0}}{\partial p} \Big] 
    = -\frac{f_{0}}{T_{\rm esc}},
\end{dmath}
where $T_{\rm esc}$ is the escape timescale.  
The above equation is same one used in \cite{vaidya_2018} to update the spectral distribution of a single macro-particle with the additional contributions related to Fermi $ \rm II$ order acceleration and the escape term. 

Note that, for relativistic flows, the convective derivative can be expressed as,
\begin{equation}
    u^{\mu}\nabla_{\mu}\equiv\gamma\left[\frac{\partial}{\partial t}+v^i\frac{\partial}{\partial x^i}\right] = \frac{d}{d\tau},
\end{equation}
where $\tau$ is the proper time.
Assuming pitch angle isotropy in momentum space ($p$), the distribution function can be written in terms of the number density of the relativistic particles as $N(p,\tau) dp = 4\pi p^{2}f_{0} dp$ with $N(p,\tau)$ being the number density of non-thermal particles with momentum between $p$ and $p+dp$. 
Accordingly Eq.~(\ref{eq:webb_1}) can be written as,
\begin{align}\label{eq:webb_2}
   \nonumber \frac{d N}{d\tau} +\frac{\partial}{\partial p}\Bigg[-N\nabla^{\mu}u_{\mu}\frac{p}{3}+\frac{\left<\dot p\right>_l}{p^2}N - D_{pp}\frac{\partial N}{\partial p} \\
    +\frac{2N D_{pp}}{p}\Bigg] = -N\nabla^{\mu}u_{\mu}- \frac{N}{T_{\rm esc}}
\end{align}
Transforming the independent variable from momentum ($p$) to Lorentz factor ($\gamma$) following $p\approx \gamma m_0 c$, with $c$ being the speed of light in vacuum and $m_0$ being the mass of the ultra relativistic cosmic ray particles, Eq.~(\ref{eq:webb_2}) can be expressed as \citep[see Eq.~11 of][]{tramacere_2011}:
\begin{equation}\label{eq:main}
  \begin{array}{lcl}
  \DS    \pd{\chi_p}{\tau}
  + \pd{}{\gamma}\left[(S+D_{A})\chi_{p}\right] &=&\DS
    \pd{}{\gamma}\left(D\pd{\chi_p}{\gamma}\right)
  -\frac{\chi_p}{T_{\rm esc}}  \\ \noalign{\medskip}
    & & \DS +Q(\gamma,\tau)\,,
 \end{array}
\end{equation}
where $\chi_{p}=N/n$, with $n$ being the number density of the fluid at the position of macro-particle, $S$ corresponds to radiative and adiabatic losses and $D_{A} = 2D/\gamma^{2}$ corresponds to the acceleration due to Fermi $\rm II$ order with $D = D_{pp}/m_{0}^{2}c^{2}$. 
We also include $Q(\gamma,\tau)$ as a source term in Eq.~(\ref{eq:main}), which accounts for particle injection process from external sources.

A numerical approach to solve Eq.~(\ref{eq:main}) without the terms on the right hand side and $D_{A}$ has been discussed in an earlier work \citep{vaidya_2018}, along with the particle energization through $1^{\rm st}$-order Fermi acceleration at shocks.
The numerical method for DSA has then recently been improved to account for the history of particle spectra by \cite{mukherjee_2021} and will be repeated here for completeness.

The improved version of the DSA routine includes a convolution of the upstream spectra to the downstream region of the shock in an instantaneous steady state manner.
In particular, as the macro-particle crosses the shock, its downstream spectra is updated as follows: 
\begin{eqnarray}\label{eq:shck_update}
\chi_{p}^{\rm down}(\gamma) \propto \int_{\gamma_{\min}}^{\gamma} \chi_{p}^{\rm up}(\gamma')G(\gamma,\gamma')\frac{d\gamma}{\gamma}
\end{eqnarray}
where, $\chi_{p}^{\rm up}(\gamma)$ is the distribution function far upstream and $\chi_{p}^{\rm down}(\gamma)$ is the steady state downstream distribution function, $G(\gamma,\gamma')=(\gamma/\gamma')^{-m+2}$, with $m=3r/(r-1)$ and $r$ is the compression ratio. 
Here, $\gamma_{\min}$ is the minimum value of Lorentz factor obtained from the upstream spectrum.  
The value of $\gamma_{\max}$, the upper-limit of the convolution, is evaluated by equating timescales due to radiative losses and various acceleration processes (i.e., DSA and STA) \citep{bottcher_2010,mimica_2012,vaidya_2018}.
Further, it is also ensured that the Larmor radius of the highest energetic lepton within a macro-particle has a radius equal to or less than one grid cell width. 
Further details are explicitly mentioned in \citep{vaidya_2018,mukherjee_2021}.

\subsection{Momentum diffusion coefficient ($D$)}
%
%

The micro-physical processes of the turbulent interaction are encapsulated in the transport coefficients of Eq.~(\ref{eq:main}). 
The mathematical form of these transport coefficients due to different interactions of cosmic ray and turbulent magnetized medium have been derived for Alfv\`enic turbulence \citep[see, for instance,][]{schlickeiser_2002, brunetti_2007, sullivan_2009}. 

In this work, we will consider STA following a 1D energy spectrum expressed as a power-law in terms of wave vector norm $|\vec{k}|=k$ with exponent $-q$, 
\begin{eqnarray}\label{eqn:turb}
    W(k)\sim k^{-q},
\end{eqnarray}
where, $W(k)$ is the turbulent energy spectrum in Fourier space. The momentum diffusion coefficient can therefore be expressed as \citep{schlickeiser_1989,sullivan_2009},
\begin{eqnarray}\label{eqn:mom_diff}
    D_{pp} \approx \beta_{A}^{2}\frac{\delta B^{2}}{B^{2}}\Big(\frac{r_{g}}{\lambda_{\max}}\Big)^{q-1} \frac{p^{2}c^{2}}{r_{g} c} \propto p^{q},
\end{eqnarray}
where $p$ is the momentum of the cosmic ray particles, $D_{pp}$ is the momentum diffusion coefficient, $\beta_{A}$ is the Alf\'ven velocity normalized to the speed of light, $B$ is the mean magnetic field, $\delta B$ its fluctuations, $r_{g}$ is the particle gyroradius and $\lambda_{\max}$ is the  maximum correlation length of the turbulent medium.

With the definitions above, the systematic acceleration timescale ($t_{A}$) for STA can be written as
\begin{eqnarray}\label{eqn:tacc}
    t_{A}\approx \beta_{A}^{-2}\frac{l}{c}.
\end{eqnarray}
where $l$ (the mean free path of the cosmic ray particle) can be expressed as
\begin{eqnarray}\label{eqn:length}
    l\approx\frac{B^{2}}{\delta B^{2}}\Big(\frac{r_{g}}{\lambda_{\max}}\Big)^{1-q} r_{g}.
\end{eqnarray}
Therefore, the acceleration timescale (Eq.~(\ref{eqn:tacc})) in terms of $\gamma$ could be expressed as,
\begin{eqnarray}\label{eqn:main}
    t_{A}\approx\frac{A^{2}}{2}\rho  c (m_{0}\gamma c^{2})^{2-q} B^{q-4} \lambda_{\max}^{q-1},
\end{eqnarray}
where, $A=B/\delta B$ defines the turbulence level whose value is set to unity for the present study \citep{sullivan_2009}. 

\subsection{Timescales}\label{sec:timescale}
%
%

The processes described in Eq.~(\ref{eq:main}) involve separate timescales due to different radiative losses and STA process. 
These timescales can be expressed in terms of the particle Lorentz factor $\gamma$ as follows:
\begin{enumerate}
    \item Radiative losses time due to Inverse Compton (IC) in Thompson limit and synchrotron radiation, $t_{L}\propto 1/\gamma$;
    \item Diffusion time due to Fermi $\rm II$ order momentum diffusion $t_{D}\propto \big(\frac{\gamma}{\gamma_{s}})^{2-q}$, for the chosen diffusion coefficient $D\propto \Big(\frac{\gamma}{\gamma_{s}}\Big)^{q}$. The value of $t_{D}$ therefore becomes a constant, $t_{D}=1/D_{0}$ with a choice of $q = 2$, where $D_{0}$ is the proportionality constant.  Here, $\gamma_{s}$ defines scale Lorentz factor which we have taken it to be unity for all the cases considered in this work;
    \item The acceleration timescale $t_{A} = t_{D}/2$, estimated from Eq.~(\ref{eq:main}) with the acceleration coefficient $D_{A}= 2D/\gamma$. 
\end{enumerate}

These considerations are of crucial importance in devising a numerical scheme for the solution of Eq.~(\ref{eq:main}), since an explicit method would demand $\Delta t < {\min}\{t_{L},\,t_{D},\, t_{A}\}$ for stability reason.
%
\section{Turbulent Particle Acceleration : Algorithm}
\label{sec:IMEX}
%
%
%
\subsection{Numerical Method}
%
%

Eq. ~(\ref{eq:main}) is a non-homogeneous, convection-diffusion like partial differential equation (PDE) with variable coefficients.
This equation combines both hyperbolic and parabolic terms.
The non-homogeneous character of the equation is attributed to the presence of the source and sink terms.

While various numerical methods for the numerical solution of Eq. (\ref{eq:main}) have been proposed \citep[see, for instance][]{chang_1970,winner_2019}, here we take a more up-to-date and refined approach based on the employment of Runge-Kutta IMplicit-EXplicit (RK-IMEX) schemes whereby the hyperbolic term of the PDE are treated using an upwind Godunov-type explicit formalism while the parabolic (diffusion) term is handled implicitly. 

Also, in order to account for the large range of values taken by the particle Lorentz factor $\gamma$, we employ a logarithmically spaced grid to provide equal resolution per decade.

To this end, we first introduce a coordinate transformation for the independent coordinate $\gamma \in [\gamma_{\min},\,\gamma_{\max}]$ in the following way,
\begin{equation}\label{eq:grid_trans}
    \xi(\gamma) =  \frac{\log(\gamma/\gamma_{\min})}
                        {\log(\gamma_{\max}/\gamma_{\min})}, 
\end{equation}
where, $\xi \in [0,1]$ is the transformed (logical) coordinate. 
Eq.~(\ref{eq:main}) is then rewritten as,

\begin{dmath}\label{eq:xi_log}
      \frac{\partial \chi}{\partial\tau} 
     + \xi'\frac{\partial }{\partial \xi} (H\chi) 
        = 
     \xi'\frac{\partial }{\partial \xi}
    \Big[D\xi'
        \frac{\partial \chi}{\partial \xi}\Big] 
        - \frac{\chi}{T_{ esc}}+Q\,
\end{dmath}
where we have dropped the subscript $p$ for ease of notation, while $\xi'$ is the Jacobian of this transformation given by Eq. (\ref{eq:grid_trans}),
\begin{equation}
    \xi' = \frac{d\xi}{d\gamma}=\frac{1}{\gamma\log(\gamma_{\max}/\gamma_{\min})}\,,
\end{equation}
while $H = S + D_A$, from Eq. (\ref{eq:main}).

In order to apply the RK-IMEX scheme, we discretize Eq.~(\ref{eq:xi_log}) on a one-dimensional mesh of $N$ points using the method of lines,

\begin{dmath}\label{eq:RK_IMEX}
    \frac{d\chi_i}{dt}={\cal A}_i + {\cal D}_i + {\cal S}_i, 
\end{dmath}
so that the original PDE becomes a system of ordinary differential equations at the nodal points $i=i_b,\,...,\,i_e$, with $N = i_e - i_b + 1$.
In Eq. (\ref{eq:RK_IMEX}), ${\cal A}_{i}$ is the advection term, ${\cal D}_{i}$ is the diffusion term and ${\cal S}_{i}$ accounts for accounts for source and sink terms.

The advection term ${\cal A}_{ i}$ is discretized in conservative fashion using the nonlinear Van Leer flux limiter scheme \citep{vanleer_1977},
\begin{equation}\label{eq:adv_part}
    {\cal A}_{i}=-\xi'_i\frac{ {\cal F}^{ \rm adv}_{i+\HALF}  - {\cal F}^{\rm adv}_{i-\HALF}}{\Delta \xi},
\end{equation}
where the advection flux follows an upwind selection rule,
\begin{equation}
    {\cal F}_{i+\HALF}^{\rm adv}=\begin{cases}
       H(\gamma_{i+\HALF}) \chi_{i+\HALF}^{L} & \quad {} \quad H(\gamma_{i+\HALF})>0 \\
       H(\gamma_{i+\HALF}) \chi_{i+\HALF}^{R} & \quad {} \quad H(\gamma_{i+\HALF})<0  \,.
    \end{cases}
\end{equation}
The left and right states $\chi^{L}_{i+\HALF}$ and $\chi^{R}_{i+\HALF}$ are constructed up to $2^{\rm nd}$-order accuracy in space using a slope limiter to prevent oscillations around extrema,
\begin{equation}
\begin{array}{l}
    \DS \chi^{L}_{i+\HALF} = \chi_i     + \frac{\delta \chi_i}{2}    , \\ \noalign{\medskip}
    \DS \chi^{R}_{i+\HALF} = \chi_{i+1} - \frac{\delta \chi_{i+1}}{2},
\end{array}
\end{equation}
with the $\Delta\chi_i$ is the harmonic mean slope limiter \citep{vanleer_1977},
\begin{equation}
    \delta \chi_{i} = \begin{cases}
   \DS  \frac{2\Delta \chi_{i+\HALF}\Delta\chi_{i-\HALF}}
         {\Delta\chi_{i+\HALF} + \Delta\chi_{i-\HALF}} 
         & \quad {\rm if } \quad \Delta \chi_{i+\HALF}\Delta \chi_{i-\HALF}>0 \\
       0 & \quad {} \quad {\rm  otherwise}
    \end{cases}
\end{equation}
where, $\Delta \chi_{i\pm\HALF}=\pm(\chi_{i\pm1}-\chi_{i})$. 
Note that this scheme is $2^{\rm nd}$-order accurate away from discontinuities and that the reconstruction step demands for 2 ghost zones beyond the active domain cells.

For the diffusion term ${\cal D}_{i}$, we also adopt a conservative formalim and choose a central differencing approach yielding $2^{\rm nd}$-order accuracy in the uniform $\xi$ grid:
\begin{equation}\label{eq:diff_part}
    {\cal D}_{i}= \xi'_i\frac{{\cal F}^{ \rm diff}_{i+\HALF} - {\cal F}^{\rm diff}_{i - \HALF}}{\Delta \xi},
\end{equation}
where, 
\begin{equation}
    {\cal F}_{i+\HALF}^{\rm diff} = \left(\xi'D(\gamma,t)\right)_{i+\HALF}
    \Bigg( \frac{\chi_{i+1}-\chi_{i}}{\Delta \xi}\Bigg),
\end{equation}
is the diffusion flux constructed following a central difference approach.

In the RK-IMEX approach, the advection is carried out explicitly while the diffusion operator and the source terms are handled implicitly.
This allows to overcome the restrictive time step limitation $\Delta t \lesssim \Delta\xi^2/(\xi'D)$ imposed by a typical explicit discretization.

We have implemented two similar approaches for the temporal integration of Eq.~(\ref{eq:RK_IMEX}) in the PLUTO code.
The first one is the Strong Stability Preserving (SSP) scheme (2,2,2) of \cite{pareschi_2005}. 

Omitting the subscript $i$ for simplicity,
\begin{equation}\label{eq:SSP222}
  \begin{array}{lcl}
  \chi^{(1)} &=&\displaystyle \chi^{(n)} + \Delta t\alpha{\cal D}^{(1)}
    \\ \noalign{\medskip}
  \chi^{ (2)} &=&\displaystyle \chi^{(n)} + \Delta t\Big[{\cal A}^{(1)} 
                            + (1-2\alpha){\cal D}^{(1)}
                            +    \alpha  {\cal D}^{(2)}\Big]       
    \\ \noalign{\medskip}
  \chi^{(n+1)} &=&\displaystyle \chi^{(n)} + \frac{\Delta t}{2}
      \Big[  {\cal A}^{(1)} + {\cal A}^{(2)}  
           + {\cal D}^{(1)} + {\cal D}^{(2)}  \Big],       
  \end{array}    
\end{equation}
where $\Delta t$ is the time-step, $\alpha = 1-1/\sqrt{2}$.

For the second approach we choose ARS(2,2,2) scheme due to \cite{ascher_1997}:
\begin{equation}\label{eq:ARS222}
  \begin{array}{lcl}
  \chi^{(1)} &=&\DS \chi^{(n)} + \Delta t\Big[\alpha{\cal A}^{(n)} 
                            +    \alpha {\cal D}^{(1)}\Big]       
    \\ \noalign{\medskip}
  \chi^{(n+1)} &=&\DS \chi^{(n)} + \frac{\Delta t}{2}
      \Big[  \delta{\cal A}^{(n)} + (1-\delta){\cal A}^{(1)} \Big] 
    \\ \noalign{\medskip}
               & & \DS + \frac{\Delta t}{2}\Big[(1-\alpha){\cal D}^{(1)} + \alpha{\cal  D}^{(n+1)}  \Big],       
  \end{array}    
\end{equation}

where, $\alpha = 1-1/\sqrt{2}$, $\delta=1-\frac{1}{2\alpha}$.

Both time-stepping methods require the inversion of two tri-diagonal matrices per step, which we perform following the Thomas algorithm \citep{press_1992}. 
In the present work, we will only show results from the SSP(2,2,2) scheme since results obtained with the ARS(2,2,2) are similar.
Furthermore, for the sake of comparison, we have also implemented the standard Chang-Cooper algorithm \citep{chang_1970, park_1996} for solving the Fokker-Planck Equation.

\subsubsection{Boundary conditions} \label{sec:boundary}
%
%

In order for our numerical method to operate correctly, boundary conditions (b.c.) must be specified in the guard (or ghost) zones for $i=i_b-1, i_b-2$ and likewise for $i=i_e+1,i_e+2$. 
Two common b.c. have been routinely employed \citep{Marcowith_2020}. 
The first one (zero-particle) is a Dirichlet b.c. requiring the value of the distribution function $\chi$ to vanish in the ghost zones.
This kind of boundary condition in solving the cosmic ray transport problem is used, for instance, by \cite{winner_2019}. 
Another boundary condition is a Neumann-like condition requiring zero-flux across the boundary interface. 
This condition has been used, for instance, by \cite{chang_1970} to solve the Fokker-Planck equation. 
The zero-flux b.c. conserves the integral of $\int \chi d\gamma$ (the analogous of particle number conservation). 
For more discussion on the boundary conditions for cosmic ray transport see \cite{park_1995}. 
Unless otherwise states, we will employ the zero-flux b.c. to ensure that without the presence of source and sink terms in Eq.~(\ref{eq:main}), the total number of particles remain conserved. 
At the implementation level, we enforce the zero-flux b.c. separately according to the implicit/explicit stage level in our RK-IMEX update:
\begin{itemize}
    \item during the implicit diffusion step we impose zero-gradient b.c.:
    \begin{equation}
       \begin{dcases}
       \begin{array}{ll}
        \chi_{i}^{\rm diff}=\chi_{i_b}^{\rm diff}  & \quad {\rm for} \quad i < i_b  
        \\ \noalign{\medskip}
        \chi_{i}^{\rm diff}=\chi_{i_e}^{\rm diff}  & \quad {\rm for} \quad i > i_e
        \end{array} 
        \end{dcases}.
    \end{equation}
   where $\chi^{\rm diff}$ is the solution array immediately before the implicit step.
   \item during the explicit hyperbolic update we impose reflective condition
   \begin{equation} \label{eq:adv_bc}
       \begin{dcases}
      \begin{array}{ll}
        \chi_{i}^{\rm adv} = -\chi_{2i_b-i-1}^{\rm adv}  & \quad {\rm for} \quad i < i_b  
        \\ \noalign{\medskip}
        \chi_{i}^{\rm adv} = -\chi_{2i_e-i+1}^{\rm adv}  & \quad {\rm for} \quad i > i_e
        \end{array}
       \end{dcases}
    \end{equation}
    together with 
    \begin{equation}
        {\cal F}_{i_b-\HALF}^{\rm adv} = {\cal F}_{i_e+\HALF}^{\rm adv} = 0 \,.
    \end{equation}
    In Eq. (\ref{eq:adv_bc}) $\chi^{\rm adv}$ represents the solution array immediately before the explicit advection step.
\end{itemize}

A third b.c. is used to assess the accuracy of our algorithm against a reference or analytical solution. 
In this case, the value of $\chi$ in the ghost zones is set to the corresponding analytical value in those zones, unless otherwise stated.

\section{Results : Code Validation Tests} \label{sec:validation}
%
%

In this section we proceed to assess the accuracy of our newly proposed algorithm.
For accuracy calculation, errors will be computed using the $L_{1}$ norm, defined as \citep{winner_2019}:

\begin{dmath}
    L_{ 1}(N)= \frac{\displaystyle\sum_{i=1}^{N}\left|\chi^{ \rm ref}_{ i}-\chi^{\rm num}_{ i}\right|\Delta\gamma_{ i}}{\displaystyle\sum_{i=1}^{N}\chi^{\rm ref}_{ i}\Delta\gamma_{ i}},
\end{dmath}
where, $N$ is the number of energy bins. 
To further ensure that the scheme accuracy is not get dominated by the spatial discretization, the increment in $N$ is compensated by the decrement in $\Delta t$ such that the ratio $N/\Delta t$ stays constant \citep{vaidya_2017}. 
In section \ref{sec:with_shock} all the tests are performed following the zero-flux boundary prescription. 
Furthermore all the simulations in this work are performed using the SSP(2,2,2) scheme with Courant number 0.4, unless otherwise specified.
%
\subsection{Simple Advection}
%
%
\begin{figure}
    \centering
    \includegraphics[scale=0.3]{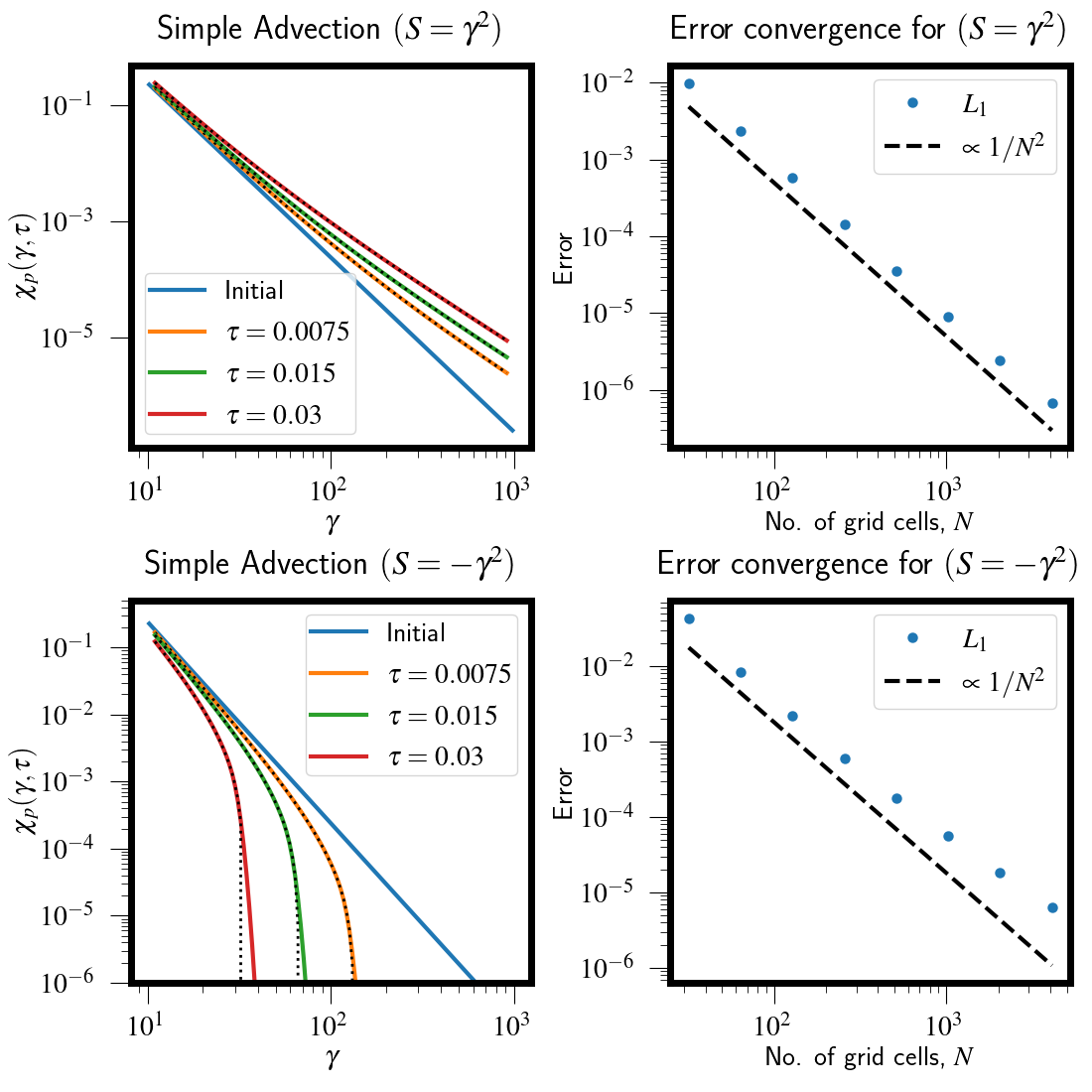}
    \caption{Evolution of the particle distribution function and their corresponding $L_{1}$ error for the simple advection following $S=\gamma^2$ (\emph{Top panel}) and $S=-\gamma^2$ (\emph{Bottom panel}) case with IMEX-SSP algorithm. \emph{Left panel:} 
     shows the numerical (solid lines) and analytical (black dotted lines) solutions at different times.
    \emph{Right panel:} L1 norm errors at different resolutions (blue dots) and $2^{\rm nd}$-order reference slope (dashed lines).}
    \label{fig:pure_adv}
\end{figure}

We start by considering a simple advection benchmark by setting $S=k\gamma^{ 2}$, $D_{A}=D=0$ in Eq.~(\ref{eq:main}). Here we consider two cases, owing to two diffrent values of $k=\pm 1$. 
The analytical solution for the case of $k=-1$ is given by \citep{kardashev_1962,Sarazin_1999}:
\begin{equation}\label{eq:adv_ana}
    \chi_{ p}=
    \begin{dcases}
    N_{ 0}\gamma^{ -s}(1-\gamma/\gamma_{\rm cut})^{ s-2}, & \,\gamma\geq\gamma_{\rm cut}\\
    0,    & \,\gamma\leq\gamma_{\rm cut}
    \end{dcases}
\end{equation}
where, $\gamma_{\rm cut} = 1/\tau$, while for $k=1$ we do not encounter such discontinuity in the result,
\begin{equation}\label{eq:adv_ana_+}
    \chi_{ p}=
    N_{ 0}\gamma^{ -s}(1+\gamma/\gamma_{\rm cut})^{ s-2}.
\end{equation}
The initial condition consists of a power-law spectrum, $\chi_{ p}(\gamma,0)=N_{ 0}\gamma^{ -s}$ with  $s=3.3$. 
For the numerical calculations, we consider the range of $\gamma\in[10,10^{ 3}]$ as our computational domain. 
We show the evolution of $\chi_p$ and the corresponding error for both values of $k$ in Fig.~\ref{fig:pure_adv}, using $128$ bins and fixed time step $\Delta \tau = 0.00375$.
The top left panel of Fig.~\ref{fig:pure_adv} shows the evolution of $\chi_{p}$ for $k=1$, while the bottom left panel depicts the same for $k=-1$.   
The solid curves represent the numerical solutions while the black dotted curves depict the analytical solution at the corresponding time.
For $k=1$, the distribution function follows the analytical results closely, while, for $k=-1$  some deviations are observed at a later stage ($\tau=0.03$) between the analytic and numerical solution, owing to the steepening of the solution (Eq.~\ref{eq:adv_ana}).
A convergence test is shown for both cases in the right panel of Fig.~\ref{fig:pure_adv} where we plot the $L_1$ error as a function of the number of bins.
Blue dots and the black dashed curve represent, respectively, the computed $L_{ 1}$ error and a reference for the $1/N^{ 2}$ slope.
For $k=1$ (top right) results converge with $2^{\rm nd}$-order accuracy for all resolutions, while for $k=-1$ (bottom right) a slight deviation from the $2^{\rm nd}$-order convergence can be observed. 
This discrepancy is attributed to the discontinuous nature of analytic solution presented in Eq.~(\ref{eq:adv_ana}).

\subsection{Simple Diffusion}
%
%

\begin{figure}
    \centering
    \includegraphics[scale=0.3]{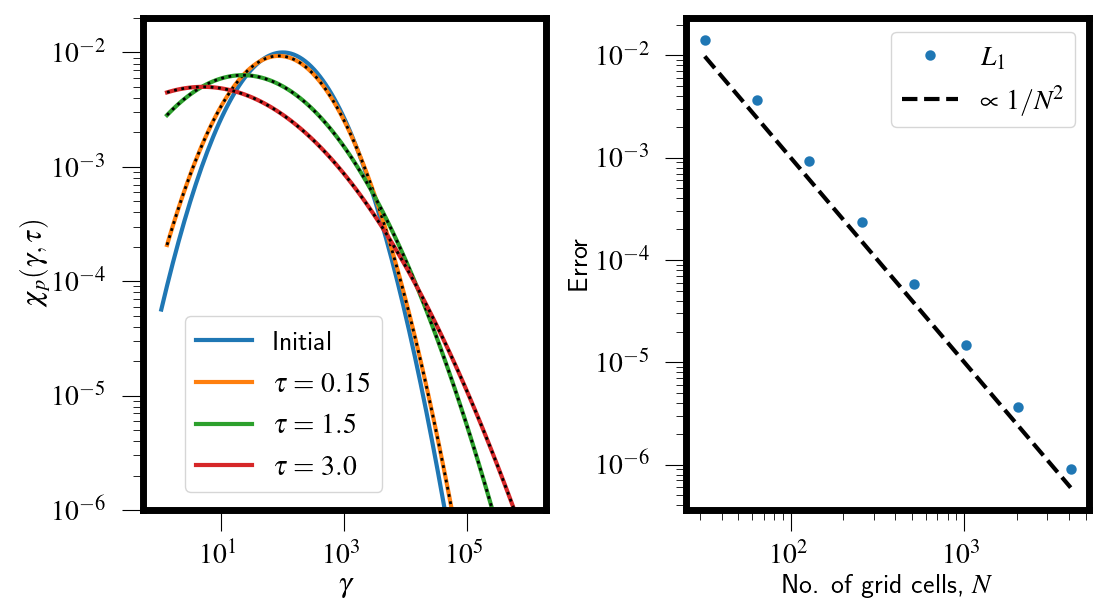}
    \caption{\emph{Left:} Simple diffusion case for different times where solid lines show the numerically computed particle distribution function and black dotted curve depicts analytical solutions. \emph{Right:} $L_{ 1}$ error convergence plot for the Simple diffusion case with IMEX-SSP algorithm.}
    \label{fig:pure_diff}
\end{figure}

Next, we solve Eq.~(\ref{eq:main}) in the case of simple diffusion where, $S=D_{ A}=0$ and $D=\gamma^{ 2}$. 
The analytical solution for this case can be written as \citep{park_1995},
\begin{equation}\label{eq:diff_ana}
    \chi_{ p} = \frac{1}{\gamma\sqrt{4\pi\tau}}\exp\Bigg\{-\frac{[\log(\gamma_{ 0}/\gamma)+\tau]^{ 2}}{4\tau}\Bigg\}
\end{equation}
We define the computational domain as $\gamma\in[1,10^{ 6}]$ and employ $128$ logarithmically spaced bins with a fixed time-step $\Delta\tau=0.0375$. 
The initial condition is given by the analytical solution (Eq.~\ref{eq:diff_ana}) at $\tau=1.0$ and $\gamma_{ 0}=100.0$. 
The results are shown in Fig.~\ref{fig:pure_diff}. 
The left panel shows the evolution of the distribution function at different times with solid (black dotted) curve representing the numerical (analytical) solution. 
In the right panel of Fig.~\ref{fig:pure_diff} the corresponding $L_{ 1}$ error is shown by varying the grid size from $32$ to $4096$ bins. 
Here $2^{\rm nd}$-order convergence is observed uniformly at all resolutions.

\subsection{Hard-sphere Equations}\label{sec:hard_sp}
%
%
The next numerical benchmark is intended to verify the correctness of our implementation when source and sink terms are present in the Fokker-Planck equation.
Additionally, we also compare our code with the standard Chang-Cooper algorithm \citep{chang_1970}. 
For this purpose, we solve the following Fokker-Planck equation
\begin{figure}
    \centering
    \includegraphics[scale=0.3]{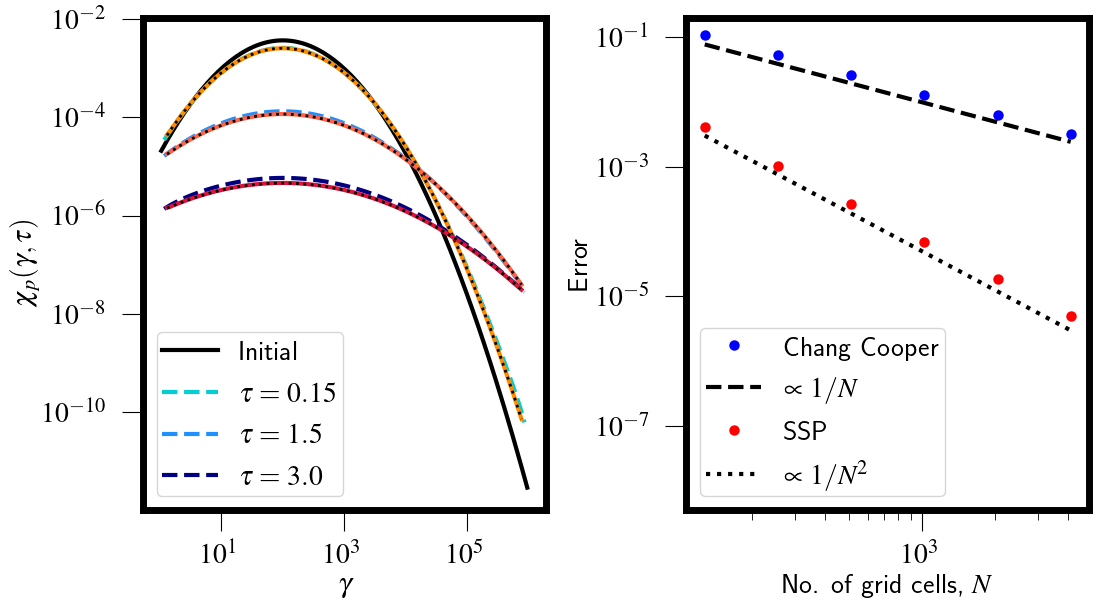}
    \caption{\emph{Left:} Evolution of the particle distribution following Eq.~(\ref{eq:FP1}) with $\theta=1$. 
    Dashed curves plot results obtained with the Chang-Cooper scheme, red curves correspond to the SSP(2,2,2) scheme.
    Different shades correspond to different times.
    Black dotted curve depicts the analytical solutions at the corresponding times.
    \emph{Right:} $L_1$-norm error convergence for both Chang-Cooper (blue dots) and SSP(2,2,2) (red dots) schemes.
    Black curves shows the reference slopes for the corresponding schemes.}
    \label{fig:chang_SSP}
\end{figure}
\begin{figure}
    \centering
    \includegraphics[scale=0.4]{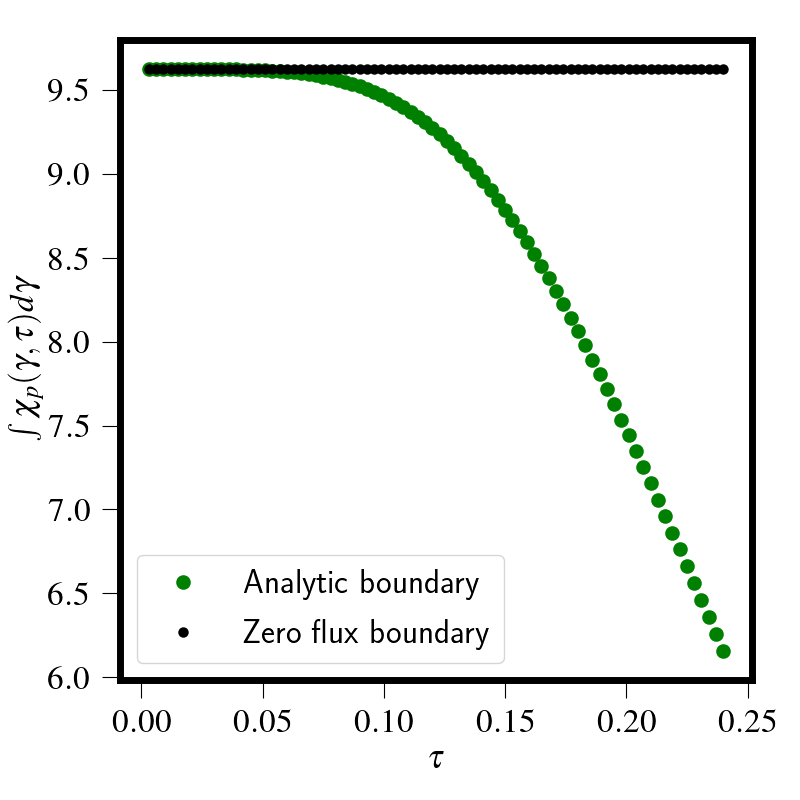}
    \caption{Time evolution of the integral $\int\chi_{ p}(\gamma,\tau)d\gamma$ is shown for the proposed boundary condition (zero flux boundary) along with the boundary condition where the value of the distribution functions in the ghost zones are computed from the analytic expression (analytic boundary).}
    \label{fig:part_num}
\end{figure}

\begin{dmath}\label{eq:FP1}
    \frac{\partial \chi_{ p}}{\partial \tau}=\frac{\partial}{\partial \gamma}\Big(\gamma^2\frac{\partial \chi_{ p}}{\partial \gamma}-\gamma \chi_{ p}(\gamma, \tau) \Big)-\theta\chi_{ p} \,.
\end{dmath}
The analytical solution of the previous equation can be written as \citep{park_1995},

\begin{equation}\label{eq:FP1_ana}
    \chi_{ p} = \frac{e^{ -\theta\tau}}{\gamma\sqrt{4\pi\tau}}\exp\Bigg\{-\frac{[\log(\gamma_{ 0}/\gamma)+2\tau]^{ 2}}{4\tau}\Bigg\} \,.
\end{equation}
For the present purpose, we take the inverse escape timescale $\theta=1$ and the initial particle distribution is obtained by setting $\tau=1.0$, $\gamma=\gamma_{0}=100.0$ in Eq.~(\ref{eq:FP1_ana}). 
The computational domain is taken as $\gamma\in[1,10^{ 6}]$ using  $128$ (log-spaced) energy bins and a fixed time step $\Delta\tau=0.0375$. 

Numerical solutions obtained via the Chang-Cooper algorithm (dashed  curves) and the SSP(2,2,2) algorithm (solid lines) are shown in the left panel of Fig.~\ref{fig:chang_SSP} at different time (colors).
The analytical solution (dotted lines) is also superposed.
The corresponding resolution study is reported in the right panel of the same figure using $L{_1}$ error.
From the plots it clearly appears that the Chang-Cooper algorithm converges at $1^{\rm st}$-order rate while the SSP(2,2,2) scheme gives full $2^{\rm nd}$-order convergence, so that even at low resolutions the latter yields an error which is already one order of magnitude smaller than the former.
At the resolution of $N = 4096$ the SSP method outperforms the Chang-Cooper scheme by more than $3$ orders of magnitude.

Notice that, although we employ a conservative discretization, particle number is not strictly conserved for this test, owing to the chosen boundary condition which allows a non-zero net flux through the endpoints of the computational domain.
In order to check particle conservation, we have therefore repeated the same test in absence of sink ($\theta=0$) and by prescribing the zero-flux b.c. (see section \ref{sec:boundary}). 
Results for the previous and current b.c. are shown in Fig.~\ref{fig:part_num}. It can be observed from the figure that while the integral due to the previous b.c (depicted by green dots), decreasing with time, the integral due to the zero-flux b.c. (depicted by black dots) remains constant. 
This validates the particle number conserving nature of the proposed boundary condition.  
%
\subsection{Log-Parabolic Nature of Particle Spectra}\label{sec:log_para}
%
%

\begin{figure}
    \centering
    \includegraphics[scale=0.32]{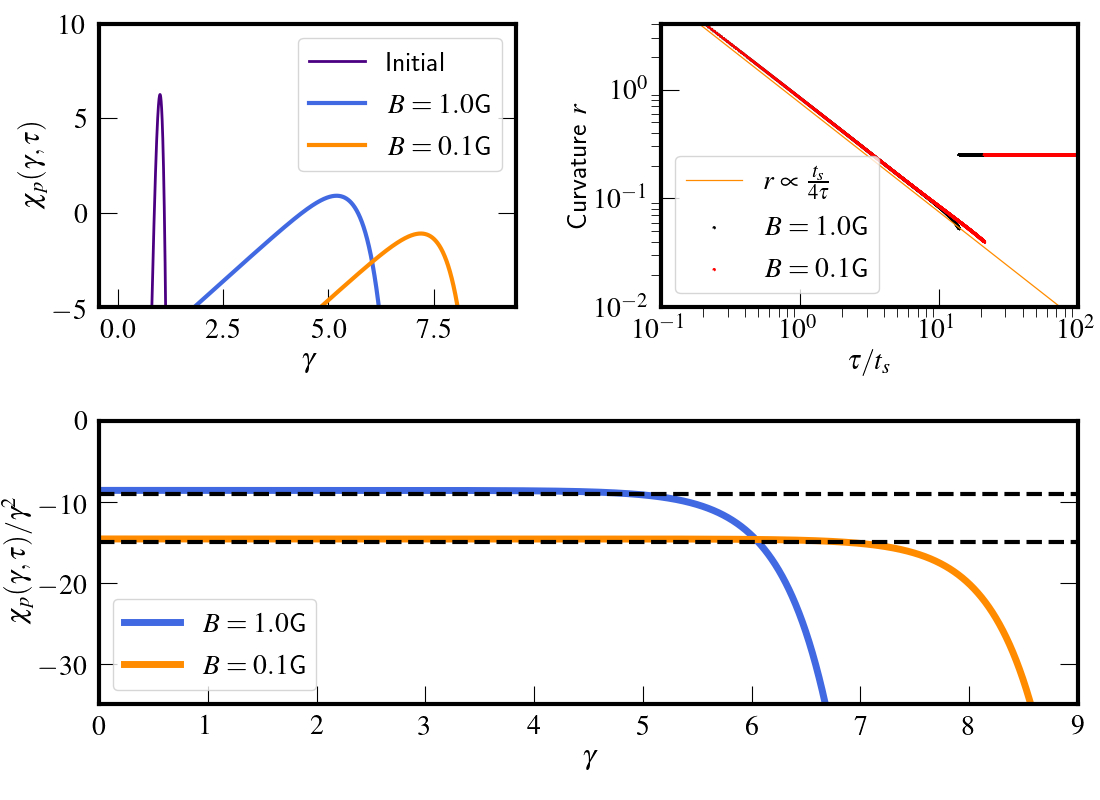}
    \caption{\emph{Top left}: evolution of the particle distribution function with turbulent acceleration and synchrotron losses with two magnetic field values.
    \emph{Top right}: evolution of the curvature of the distribution function fitted with a log-normal density profile (Eq.~\ref{eq:fit}). Analytic solution is shown in solid orange line. 
    \emph{Bottom panel}: $\chi_{ p}(\gamma,\tau)/\gamma^2$ as a function of $\gamma$ at steady state ($\tau = 30\, t_s$), in agreement with Eq.~\ref{eq:steady_q_2}. The plot shows the increase as $\gamma^2$ (black dashed lines) followed by an exponential cut-off.}
    \label{fig:dist_evolve}
\end{figure}

It has been shown \citep{massaro_2004,massaro_2006} that the hump structure in the spectral energy distribution (SED) of blazars could be described with a log-parabolic curve and this log-parabolicity is speculated to have originated from STA \citep{tramacere_2011}. 
Here we validate the log-parabolic nature of the particle distribution due to STA which consequently translates to log-parabolic nature of observed SED. 
In particular, we numerically solve the transport equation (\ref{eq:main}), in its conservative form (without source and sink terms) using the zero-flux boundary prescription, for STA including synchrotron losses.
We choose our grid as $1.0\leq\gamma\leq 10^{ 9}$ with $5000$ computational bins and $\Delta \tau=0.003$ with the following transport coefficients,
\begin{equation}\label{eq:tramacer_transport}
S = -C_0\gamma^2B^2 \,,\quad
D = D_0\gamma^2\,,\quad
D_A = \frac{2D}{\gamma} \,,
\end{equation}
where $C_{0}=1.28 \times 10^{ -9}$, $D_{0}= 10^{-4}$ sec$^{-1}$ is the diffusion constant. 
We employ $1/D_{0}$ as our unit time ($t_{s}$).

Here, we consider the one-zone model for the blazar emission \citep{tramacere_2011} where the geometry of the acceleration region is taken as spherical with radius $R=5\times10^{13}$ cm threaded by a magnetic field $B_{\rm mag}$. In this region, the acceleration is accompanied by the radiative losses. Moreover, in order to solve Eq.~(\ref{eq:main}) we consider a mono-energetic initial distribution $\chi_p$ corresponding to a total power $L_{\rm inj}=10^{39}$ erg/sec, where
\begin{equation}
L_{\rm inj}=N_{\rm part}\frac{4}{3}\pi R^3 \int \gamma m_e c^2\delta(\gamma-\gamma_{\rm inj})d\gamma,
\end{equation}
where, $N_{part}$ is the total number of particles injected per unit volume  and $\gamma_{\rm inj}=10.0$. 
The Dirac delta is approximated with a Gaussian distribution with $\sigma = 0.5$ and $\mu = 10$ and it is shown by the purple solid line in left panel of Fig~\ref{fig:dist_evolve}. 
Furthermore, 
Eq.~(\ref{eq:main}) is solved by adopting two different magnetic field values $B_{\rm mag} = 1$G, $0.1$G and the corresponding distribution of $\chi_p$ for time $\tau=30\,t_{s}$ is shown in the top left panel of Fig.~\ref{fig:dist_evolve}.

The numerical solution is shown in the top left panel of Fig.~\ref{fig:dist_evolve} for different magnetic field strengths.
We point out that the steady-state distribution is expected to have an ultra-relativistic Maxwellian form as described in Eq. (\ref{eq:steady_q_2}) in Appendix~\ref{sec:analytic}. 
This is confirmed in the bottom panel of Fig.~\ref{fig:dist_evolve} where we plot $\chi_{p}/\gamma^{ 2}$ as a function of $\gamma$, showing that our results correctly reproduce the $\gamma^2$-dependence of the spectrum.

Also, in order to quantify the effects of acceleration and radiative losses on the spectral evolution, we estimate the curvature of the distribution function. 
The curvature is measured by finding the peak value of the distribution function at each time-step which is also the point at which $t_{L}=t_{A}$ \citep[][see also Sec. \ref{sec:timescale}]{katarzynski_2006} and subsequently fitting a log-normal curve through $10$ points centered around $\gamma_{c}$ (the energy at which the maximum occurs).
The curvature is then taken as the inverse of the variance of the best fit. 
In particular, we adopt the fitting curve \citep{kardashev_1962} as follows: 
\begin{equation}{\label{eq:fit}}
 \chi_{\rm fit} =\frac{A}{\gamma \sigma} \exp{\Big\{-\frac{(\log(\gamma)-\mu-\sigma^2)^2}{4\sigma^2}\Big\}} \,,
\end{equation}
with curvature parameter defined as $r=1/(4\sigma^2)$. 
The fitting curve is a solution to the Fermi $\rm II$ order transport equation (Eq.~ \ref{eq:main} with $S=0, D=\gamma^{2}$ and $D_{A}=2D/\gamma$ without sources and sinks) when $\sigma^2=\tau$, therefore the evolution of the curvature $r$ goes as $\sim 1/(4\tau)$.
In the top right panel of Fig.~\ref{fig:dist_evolve} we compare $r$ in the acceleration region (yellow solid line) with $r$ numerically calculated by fitting Eq.~(\ref{eq:fit}) with the particle distribution, at each time, for different $B_{\rm mag}$ values (red and black dotted lines) .

Our results show that the fitted curvature initially decays with time as $r\propto t_{s}/4\tau$, following a trend of curvature in the acceleration region, and then a sudden jump of the curvature to the steady value of $r = 0.25$ can be observed. 
The results therefore confirm that, during the earlier stages,  STA dominates the evolution of the particle distribution function and, later, that steady state is reached much faster for stronger magnetic fields, as confirmed by the curvature evolution (black dots in the top right plot of Fig.~\ref{fig:dist_evolve}).

Summarizing, the numerical benchmarks proposed in this section validate our implementation and demonstrate that the proposed SSP(2,2,2) scheme is fully conservative and it provides full $2^{\rm nd}$-order accuracy, in contrast to  its predecessors \citep[i.e.][]{chang_1970, winner_2019} with typical $1^{\rm st}$-order accuracy. 

%
 \section{Effect of Turbulent acceleration in presence of Shocks} 
\label{sec:with_shock}
%
%

In this section, we describe the effect of STA on particle spectra in presence of shock. 
In particular, we consider several test situations where the equations of classical or relativistic MHD are solved using the PLUTO code \citep{mignone_2007} along with Lagrangian particles to model the non-thermal emission \citep{vaidya_2018, mukherjee_2021} in presence of DSA and radiative losses. 
To study the effects of STA, the newly developed algorithm (see section~\ref{sec:IMEX}) has been incorporated into the Lagrangian framework. 
The effects of DSA and STA on particle spectra and subsequent non-thermal emission signatures are compared for various test situations and discussed in the following.

\subsection{Non-relativistic MHD Planar shock}
\label{sec:planar}
%
%
Here we perform a simulation of a non-relativistic MHD planar shock interacting with a single macro-particle in a turbulent medium.
We solve the 2D ideal MHD equations with adiabatic equation of state on a Cartesian grid $x\in [0,40]$ and $y \in[0,2]$ using $1024 \times 128$ grid zones.
Initially, we place a shock wave at $x=1$ which moves towards the increasing $x$ direction.
The upstream density and pressure, $\rho_u$ and $P_u$, are taken as  $1$ and $10^{-4}$, respectively, in dimensionless units. 
A random density perturbation is added to simulate a non-homogeneous upstream medium.
The magnetic field is defined as $\vec{B}=B_{ 0}(\cos{\theta},\sin{\theta})$, where $\theta$ (the obliquity) is the angle between $\vec{B}$ and the direction of shock normal. 
For our purpose, we have considered $\theta = 30^{\circ}$ while $B_{0}$ is computed from the plasma beta, $\beta = 10^{ 2} = 2P_u/B_0^2$.

The physical units adopted for this test are: length  $\hat{L}_0=100\,{\rm pc}$, density $\hat{\rho}_0 = 10^{-2}\,{\rm amu}$ while the unit velocity is taken to be the speed of light $c$. 
With this choice, pressure will be given in units of $\hat{P}_0 = 1.5\times10^{ -5}\,{\rm dyne/cm}^{ 2}$, magnetic field in units of $\hat{B}_0=1.4\times 10^{ -2}\,{\rm G}$ and time in units of $\hat{\tau}_0=326.4\,{\rm yrs}$. 

The particle is initially located at $(x,y)\equiv(1.5,1.0)$ with an energy distribution following a steep decreasing power-law profile with index $9$. 
The grid ranges in $10 \le\gamma\le 10^{10}$ using $128$ (log-spaced) bins.
The particle spectrum (Eq.~\ref{eq:main}) is evolved accounting for synchrotron, inverse-Compton and adiabatic losses along with the diffusion effect, modelled following the STA timescale (Eq.~ \ref{eqn:main}). 
Additionally, the effect of shock is captured via the steady state update convolution, Eq.~(\ref{eq:shck_update}). 
We also vary the index $q$ for various turbulent spectra $W(k)\propto k^{-q}$ in three different scenarios: 
a) with only STA and no shock, b) both shock and STA and c) both shock and STA with the latter active only in the downstream region.
The value of $\lambda_{\max}$ is taken to be $\hat{L}_0/10^5$ for all the simulations.

\begin{figure*}
    \centering
    \includegraphics[scale=0.5]{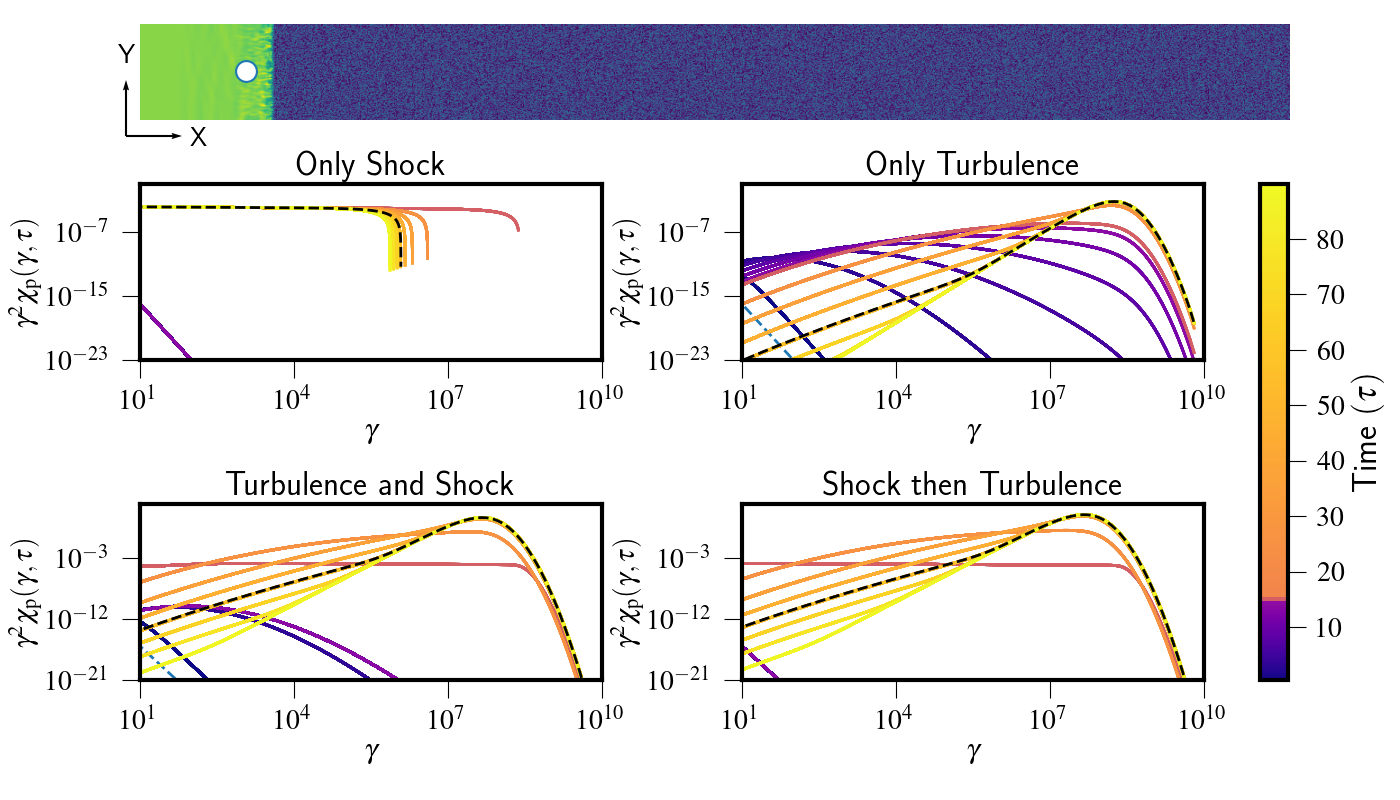}
    \caption{\emph{Top section}: Density map of a fluid with a lagrangian particle (shown in white dot). The upstream region is shown in blue, and the downstream region is shown in green. \emph{Bottom section}: Particle spectra in various scenarios with $q=2$ turbulence spectrum. Particle spectra \emph{Middle left}: For the case of only DSA with a compression ratio of $3.89$ and various losses. \emph{Middle right}: In a turbulent medium with various losses but no shock. \emph{Bottom left}: With the both shock of same compression ratio, turbulence and various losses. \emph{Bottom right}: For turbulence present only at the downstream region. The black dashed curve shows the particle energy spectrum for the time when the density map snapshot is taken.}
    \label{fig:q_2}
\end{figure*}

\begin{figure}
    \centering
    \includegraphics[scale=0.3]{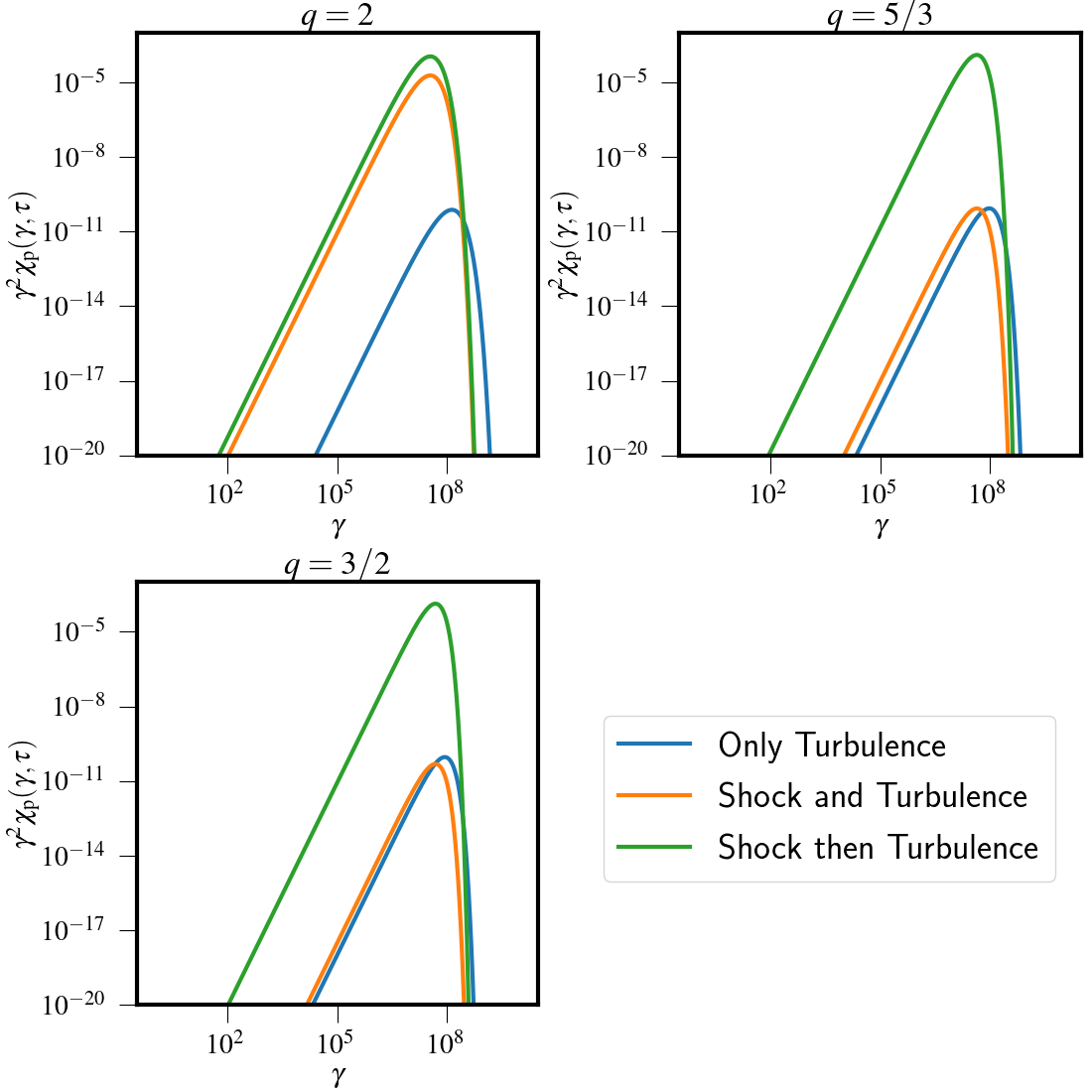}
    \caption{Steady-state particle distribution with shock and turbulence acceleration for various turbulence spectra. \emph{Left}: For $q=2$, \emph{Middle}: for $q=5/3$ and \emph{Right}: for $q=3/2$. The solid blue line depicts the case of turbulent acceleration without shock; the orange line describes the case of shock and turbulence acceleration considering both regions ahead and behind of shock are turbulent, and the green line also describes the shock and turbulence acceleration scenario where only the post-shock region is turbulent.}
    \label{fig:steady}
\end{figure}

\begin{figure}
    \centering
    \includegraphics[scale=0.3]{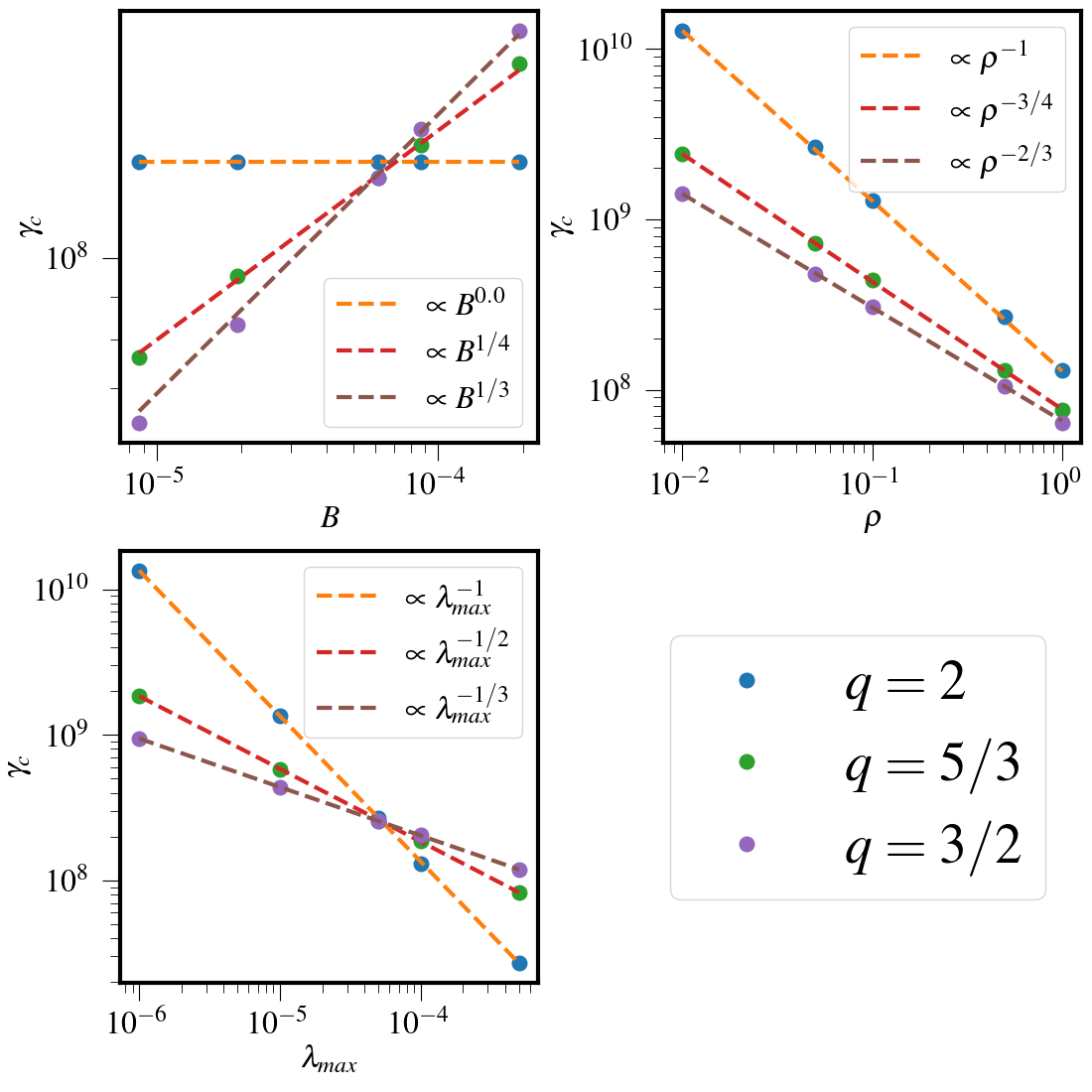}
    \caption{Dependence of $\gamma_{ c}$ on various parameters for turbulent acceleration. \emph{Left}: Dependence of $\gamma_{ c}$ on various $B$ field, \emph{Middle}: Dependence of $\gamma_{ c}$ on various $\rho$ values and \emph{Right}: Dependence of $\gamma_{ c}$ on various values of $\lambda_{\max}$. Data point from corresponding simulations are shown as dots and the result from analytic calculations (see Eq.~(\ref{eq:gamma_max})) is shown with a dashed line for reference. }
    \label{fig:gamma_All_fit}
\end{figure}

The result in the case of weak turbulence ($q=2$) is shown in Fig.~\ref{fig:q_2} where $t_A$ (see section \ref{sec:timescale}) is independent of $\gamma$. 
The top panel shows the Lagrangian particle position on top of the background gas density distribution at $t = 56.13$.
The evolution of the particle energy spectra with various radiative losses and different acceleration scenarios are shown in the bottom four panels using different colors (as indicated by the colorbar).
The upper plot depicts the evolution of the particle spectra for the situation when only DSA is effective. 
As the shock hits the particle, the spectra becomes flatter and  radiative and adiabatic losses give rise to a cut-off that gradually shifts from larger values of $\gamma$ to lower values.

The evolution of the particle spectra due to STA alone is shown in the corresponding right panel. 
The spectra is now considerably different when compared to the previous case since, owing to turbulence and losses, particle energization occurs continuously rather than just when crossing the shock.
The spectra evolves towards the typical steady state of the ultra-relativistic Maxwellian, as observed in \S\ref{sec:log_para}, with a peak value $\gamma_c \sim 10^8$ when $t_A=t_L$. 
We also notice that the high energy cut-off does not ever decreases to lower values of $\gamma$ (as for the pure DSA) but, rather, it settles into a steady state as the result of mutual compensation between losses and STA.

In the bottom left plot, we show the evolution of the energy spectrum in the presence of both shock and STA. 
Both the upstream and the downstream are turbulent.
In this scenario, the distribution function becomes harder than the initial one owing to the presence of upstream turbulence. 
The height of the spectrum now considerably increases if compared to the previous two cases. 
Such an increase is primarily due to the sub-grid modeling adopted at the shock front: the particle enters the shock with a pre-accelerated spectrum and eventually ends up in the downstream region with a different steady state (when compared to the STA alone case).

Finally, the particle energy evolution for the case in which STA is active only in the downstream region is shown in bottom right panel.
As expected, the particle distribution does not significantly change until the particle crosses the shock and then enters in the downstream region where turbulence is active. 
Here steady state is attained due to STA.
In this sense, the evolution resembles the previous case.

Further notice that, for all the cases but the pure DSA one, the particle distribution functions eventually seem to achieve steady states of similar kind.
This is expected as the predicted steady state spectrum depends on the functional form of the transport coefficients which are not affected by the presence of the shock.

\subsubsection{Effect of turbulence on evolution of particle spectra}
\label{sec:turb_other}
%
%

Additionally, in Fig.~\ref{fig:steady} we compare the particle steady-state distribution for turbulent spectra with $q=5/3$ (middle), and with $q=3/2$ (right) with that obtained for $q=2$ (left). 

The main difference between the acceleration scenario for turbulent spectrum with $q=2$, on one side, and $q=5/3$ or $q=3/2$, on the other, is that the latter achieve steady state more rapidly because of the dependence of $t_{A}$ on $\gamma$. 

Furthermore, the steady-state spectra for $q=5/3,3/2$ in the case of shock and STA are not significantly different from the ones computed with STA alone (see blue and orange solid line in the middle and right plot of Fig.~\ref{fig:steady}). 
Owing to the smaller acceleration timescale, in fact, the spectra for $q=5/3,3/2$ approach the steady state only when the particle arrives in the upstream region making the shock injection less effective (see section \ref{sec:dis}) compared to the $q=2$ case.
However, for the case where turbulence is present only in the downstream region, shock injection can clearly be observed (solid green line in Fig.~\ref{fig:steady}) as no significant turbulent energization took place in the upstream region.

Additionally, we analyze the behaviour of $\gamma_c$, with various values of $B_0$, $\rho_u$ and $\lambda_{\max}$. 
Analytically the value of $\gamma_{ c}$ can be calculated by equating $t_{ A}$ to $t_{ L}$ and yielding
\begin{equation}\label{eq:gamma_max}
    \gamma_{ c}=\left\{2\times 10^{ 3}\times\frac{\Big(\frac{eB\lambda_{\max}}{m_{ e} c^{ 2}}\Big)^{ 2-q}}{\rho\lambda_{\max}}\right\}^{\frac{1}{3-q}}
\end{equation}
Plots of $\gamma_{ c}$ computed from simulation data with different values of $B$ , $\rho$ and $\lambda_{\max}$ are compared in Fig.~\ref{fig:gamma_All_fit} toghether with the analytic form (Eq.~\ref{eq:gamma_max}).
We observe a good correspondence between the results.

\subsubsection{Interplay of DSA and STA}
%
%

\begin{figure}
    \centering
    \includegraphics[scale=0.35]{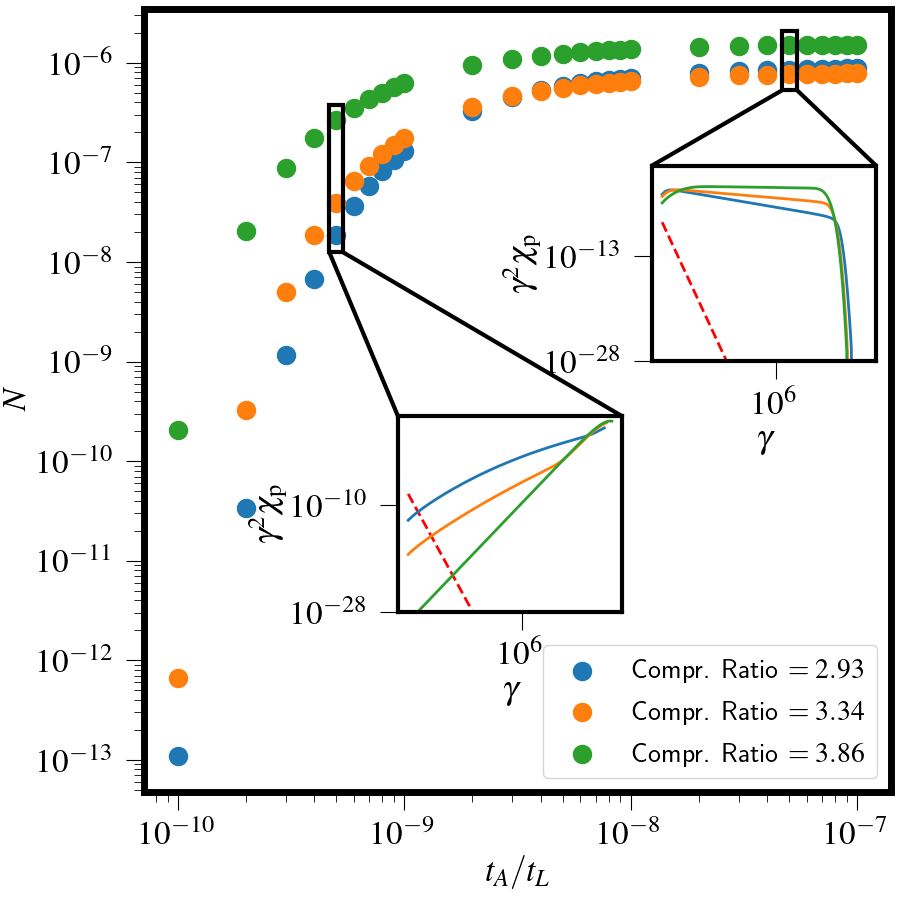}
    \caption{Dependence of shock injection on the upstream spectrum for various shock compression ratio with $\beta = 100.0$. The obliquity is made fixed at $30^{\circ}$. In the inset the downstream distribution function is shown for two different values of $t_{ A}/t_{ L}$.}
    \label{fig:compres}
\end{figure}

In the previous section we found that the shock acceleration depends on the upstream spectrum. With this motivation here we try to analyze the impact of STA on particle shock energization by modulating the acceleration timescale $t_A$ and display its effect on the shock injection with different compression ratios.
Moreover, we define the value of $t_A$ in terms of $t_{L}$ at $\gamma = 1.0$ and for each choice of $t_A$, we perform the simulation up to time $\tau = 100\,\hat{\tau}_0$.
Owing to the conserving nature of the boundary condition, the number of micro-particles in a macro-particle remains same once the shock takes place, thus by calculating the number of micro-particles after shock we estimate the effect of shock injection when STA is in process. 
The variation of total number of particles after shock is shown with ratio $t_{ A}/t_{ L}$ at $\gamma = 1.0$ for different shock compression ratio in Fig.~\ref{fig:compres} with a fixed magnetic field calculated using $\beta\,=\,100.0$. 
Further, the corresponding particle spectra at $\tau=100\,\hat{\tau}_{0}$ is plotted for two values of the ratio and is shown in the inset of Fig.~\ref{fig:compres}.  

When $t_{ A}$ is much less than $t_{ L}$ at $\gamma = 1.0$ (or the ratio $t_{ A}/t_{ L}$ is small) the particle spectrum reaches the log-parabolic steady-state (see section.~\ref{sec:planar}), before shock hits the particle. making the shock injection less effective.
On the other hand when the ratio $t_{ A}/t_{ L}$ is comparatively high, one observe very minute effect of STA on the particle distribution in the upstream making the shock injection very effective for this case. 
Furthermore, notice that for any value of $t_{ A}/t_{ L}$ shock with higher compression ratio injects more number of particles than the lower ones. 
Also from the distribution functions shown in the inset, for two different values of $t_{ A}/t_{ L}$, it can be observed that the spectra that were hit by strong shock (high compression ratio) reach to the steady state much faster compared with the spectra hit by moderate shock (moderate compression ratio).
Moreover, the decrement of the $\gamma_{ c}$ (see section \ref{sec:turb_other}) with increasing $t_{ A}/t_{ L}$ could also be seen.   
Additionally, the number could be seen to achieve a steady state, around $N \sim 10^{-6}$, at the higher values of $t_{ A}/t_{ L}$ implies an upper bound of the particle injection at the shock for different compression ratios. 

In summary, we observe that the effect of shock injection on the particle distribution function depends on the nature of the upstream particle distribution spectra. 
If the timescale of the STA in the upstream region is such that the particle distribution converges to steady-state spectra before the DSA could take place, the effect of shock injection becomes minimal. 
However, if in the upstream region the particle spectra do not reach the steady-state before the shock hits the particle, then a considerable effect of shock injection on particle spectra could be seen. 
This analysis spanning a wide parameter base, therefore showcases the interplay of these two particle acceleration processes.
%
\subsection{Relativistic Blast Wave}
%
%

\begin{figure*}
    \centering
    \includegraphics[scale=0.45]{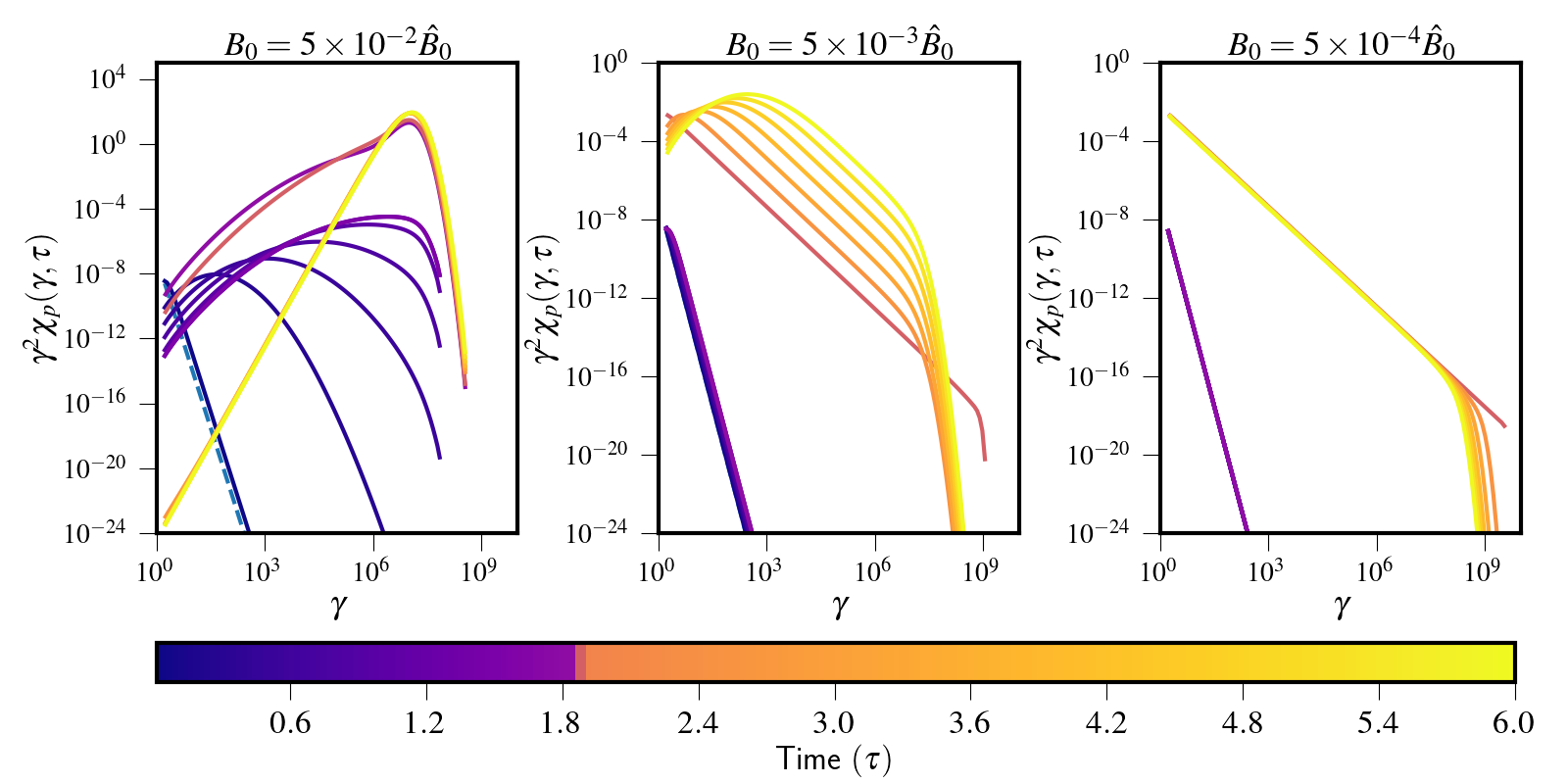}
    \caption{Temporal evolution of particle distribution of a Lagrangian particle in a turbulent medium for relativistic blast wave with different $B$ fields. The turbulent spectrum is taken as $\propto k^{ -2}$, so the value of $q$ is $2$ and the value of $\lambda_{\max}=\hat{L}_{ 0}/10$. \emph{Left}: Corresponds to $B_0=5\times10^{ -2}\hat{B}_{ 0}$, \emph{Middle}: Depicts the evolution of the particle distribution for $B_0=5\times10^{ -3}\hat{B}_{ 0}$ and \emph{Right}: Corresponds to the evolution for $B_0=5\times 10^{ -4}\hat{B}_{ 0}$. Dashed blue line corresponds to the initial distribution function which is $\propto \gamma^{ -9}$.}
    \label{fig:blast_wave}
\end{figure*}

\begin{figure}
    \centering
    \includegraphics[scale=0.42]{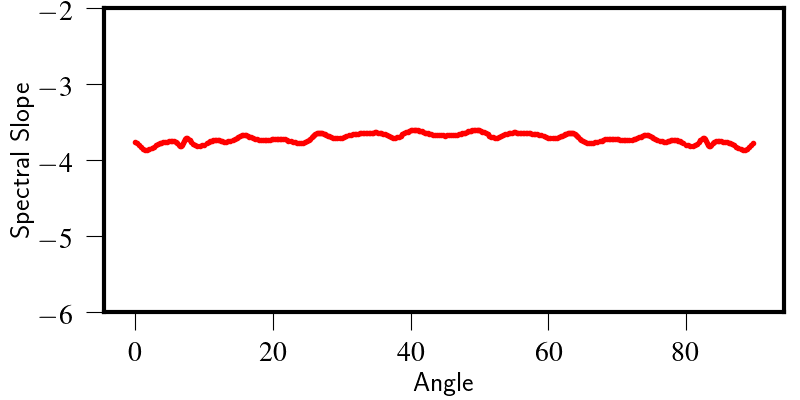}
    \caption{Spectral slope distribution of particles initially placed at different angle ($\phi$) at the final time ($\tau=6$) with $B_0=5\times10^{-4}\hat{B}_0$ for the relativistic blast wave test.}
    \label{fig:spec_dist_blast}
\end{figure}

Here we focus on the impact of a relativistic blast wave on the evolution of the spectral distribution in the presence of both shock and turbulence. 
Due to the underlying symmetry of the problem we choose a single quadrant with $512^{ 2}$ Cartesian computational zones with $x,y\in[0,6]$.
The initial condition consists of an over-pressurized central region of circular radius $0.8\hat{L}_0$ filled with pressure and density $\{P_{ c},\rho_{ c}\}=\{1,1\}$ surrounded by a uniform medium with $\{P_{e},\rho_{e}\} =\{3\times 10^{ -5},10^{ -2}\}$.
The magnetic field is taken perpendicular to the $\{x,y\}$ plane, $\vec{B}=B_{ 0}\hat{z}$ as in \cite{vaidya_2018}.
The boundary condition is set to be reflecting at $x=y=0$ and outflow elsewhere. 
We initially place $360$ Lagrangian macro-particles uniformly over $0<\phi<\pi/2$ at the radius of $\sqrt{x^2 + y^2} = 2$.
Physical units are chosen such that $\hat{L}_{0}=10\,{\rm pc}$, $\hat{\rho}_{0}=0.01\,{\rm amu}$, $\hat{P}_{0}=1.5\times10^{ -5}\,{\rm dyne/cm}^2$, $\hat{v}_{0}=c$, $\hat{B}_{0}=1.37\times10^{-2}\,{\rm G}$ and  $\hat{\tau}_{0}=32.64\,{\rm yrs}$. 
The initial distribution function for each macro-particle is taken to be a steep decreasing power-law profile with index $9$ covering a range in Lorentz factor $\gamma\in\{1,10^{8}\}$ discretized using $128$ bins.
Similar to the MHD planar shock test (section~\ref{sec:planar}), the diffusion coefficient is modelled following the acceleration timescale and the losses are modelled following the synchrotron, Inverse-Compton and adiabatic loss processes. 

The evolution of the particle distribution for a macro-particle initially placed at $65^{\circ}$, for $q=2$, is shown in Fig.~\ref{fig:blast_wave}, where the particle evolution is shown for $3$ different magnetic fields: $B_{ 0}=5\times10^{ -2}$ (left panel), $B_{ 0}=5\times10^{ -3}$ (middle panel) and $B_{ 0}=5\times10^{ -4}$ (right panel). 
Furthermore, in all three cases the value of $\lambda_{\max}=\hat{L}_{0}/10$.

For the case with strongest magnetic field, the particle distribution initially evolves due to STA and, after crossing the shock, a steady-state ultra-relativistic Maxwellian-like spectral distribution can be seen to emerge eventually with a sharp cut-off beyond $\gamma_{c} \sim 10^8$.
On the contrary, for the weakest magnetic field case, the spectral evolution shows distinct signatures of DSA only.
Indeed, STA signature can hardly be observed as the timescale obeys $t_{A}\propto B^{-2}$ (see Eq.~\ref{eqn:main}), thus very large for the simulation time.
In this case, the initial steep spectra is accelerated and the spectral slope is flattened and cooling due to synchrotron and IC emission is evident from the cut-off. 
Moreover, it should be noted that the particle can be energized beyond $\gamma > 10^9$. 
For the intermediate case, we observe effects of both shock and STA in shaping the particle spectra.

Additionally, we quantified grid orientation effects by estimating the  slope of the distribution functions for each macro-particle as a function of their initial angular positions. 
This is shown, at time $\tau=6$ for $B_0=5\times10^{-4}\hat{B}_0$, in Fig.~\ref{fig:spec_dist_blast}. 
The final slope for all the macro-particles approximately fall in the same range ($\approx-4$) with additional variations due to discretization error ($\sim2\%$).
Therefore all macro-particles will have similar spectral distribution as shown for the typical macro-particle in Fig.~\ref{fig:blast_wave}, apart from the minor variations due to discretization error.
%
\subsection{Relativistic Magneto-hydrodynamic Jet} \label{sec:jet}
%
%
\begin{figure}
    \centering
    \includegraphics[scale=0.45]{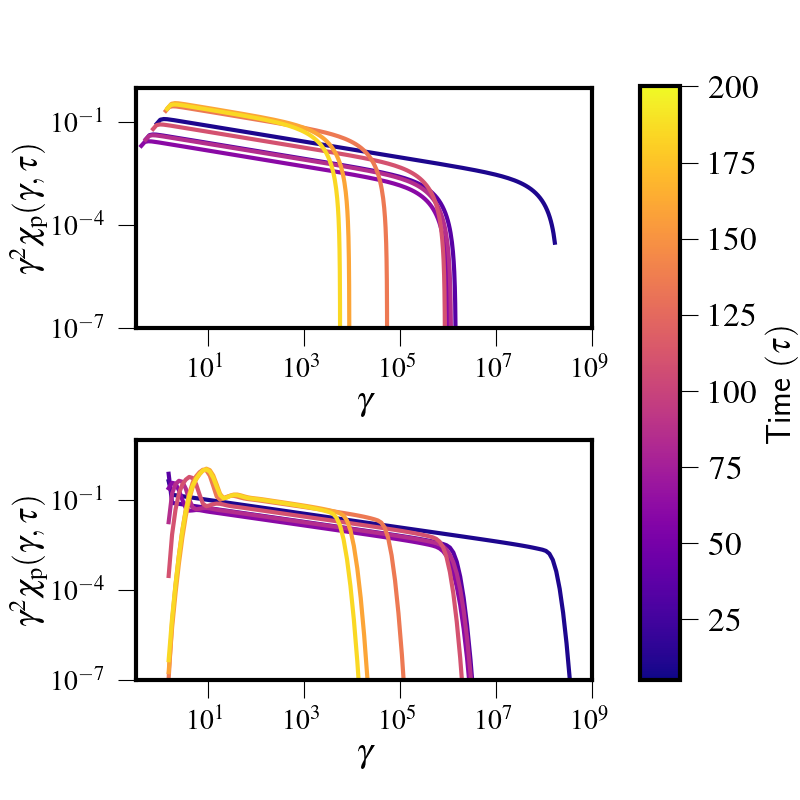}
    \caption{Temporal evolution of the spectrum of a Lagrangian particle which has gone through shock atleast once, in the RMHD Jet. \emph{Top: } For the case of only DSA \emph{Bottom: } For the case with STA along with DSA.}
    \label{fig:jet_spec}
\end{figure}

\begin{figure}
    \centering
    \includegraphics[scale=0.25]{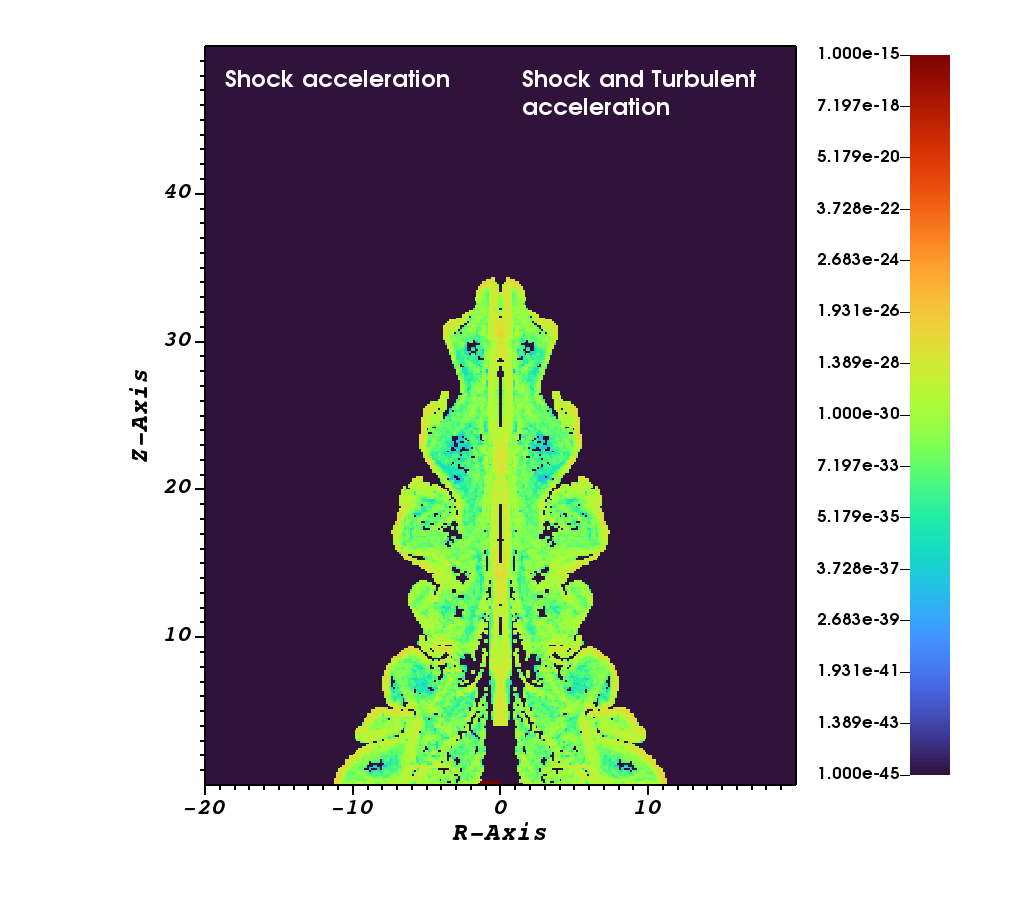}
    \vskip-2ex
    \caption{Comparison between the emission from turbulence and DSA and only DSA for radio frequency, $1.4\,GHz$ at time $\tau=200$.
    Notice that the radial coordinate has been mirrored in the left plot.}
    \label{fig:rad_1_4}
\end{figure}

\begin{figure}
    \centering
    \includegraphics[scale=0.25]{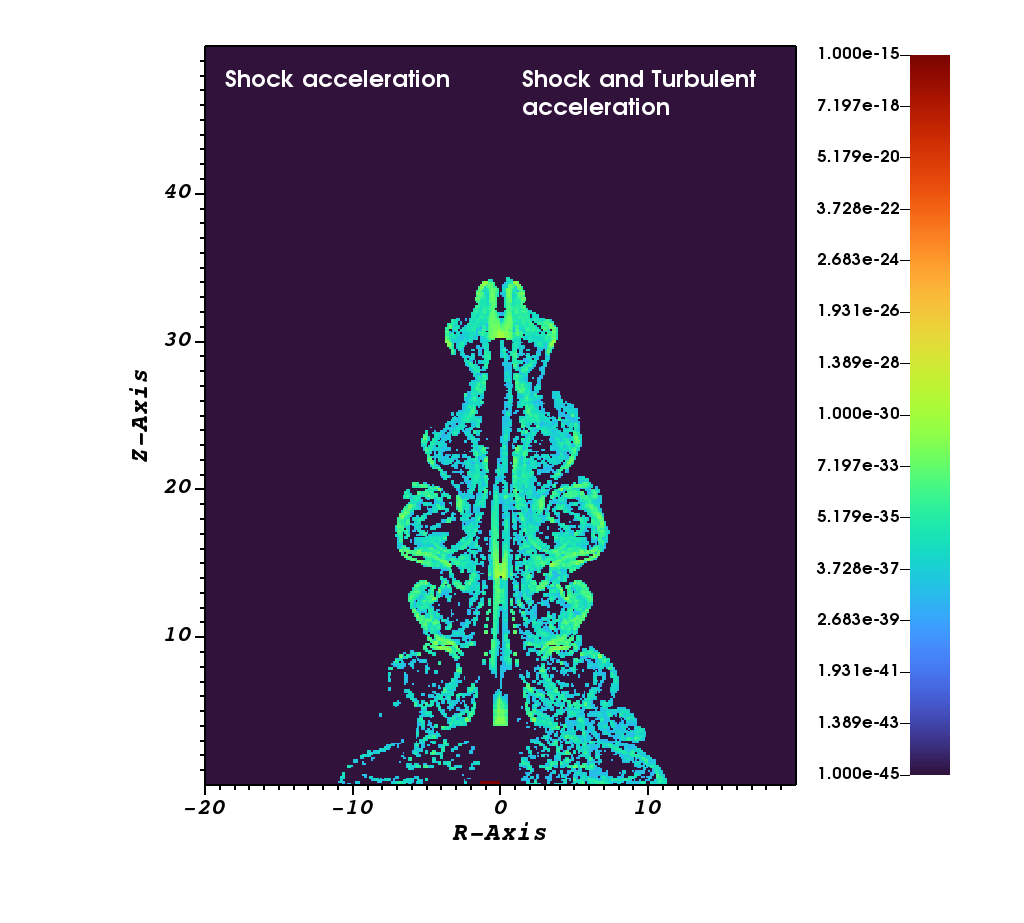}
    \vskip-2ex
    \caption{Same as Fig. \ref{fig:rad_1_4} but for optical blue light frequency $6.59\times10^{ 5}\,GHz$ at time $\tau=200$.}
    \label{fig:optical}
\end{figure}

\begin{figure}
    \centering
    \includegraphics[scale=0.25]{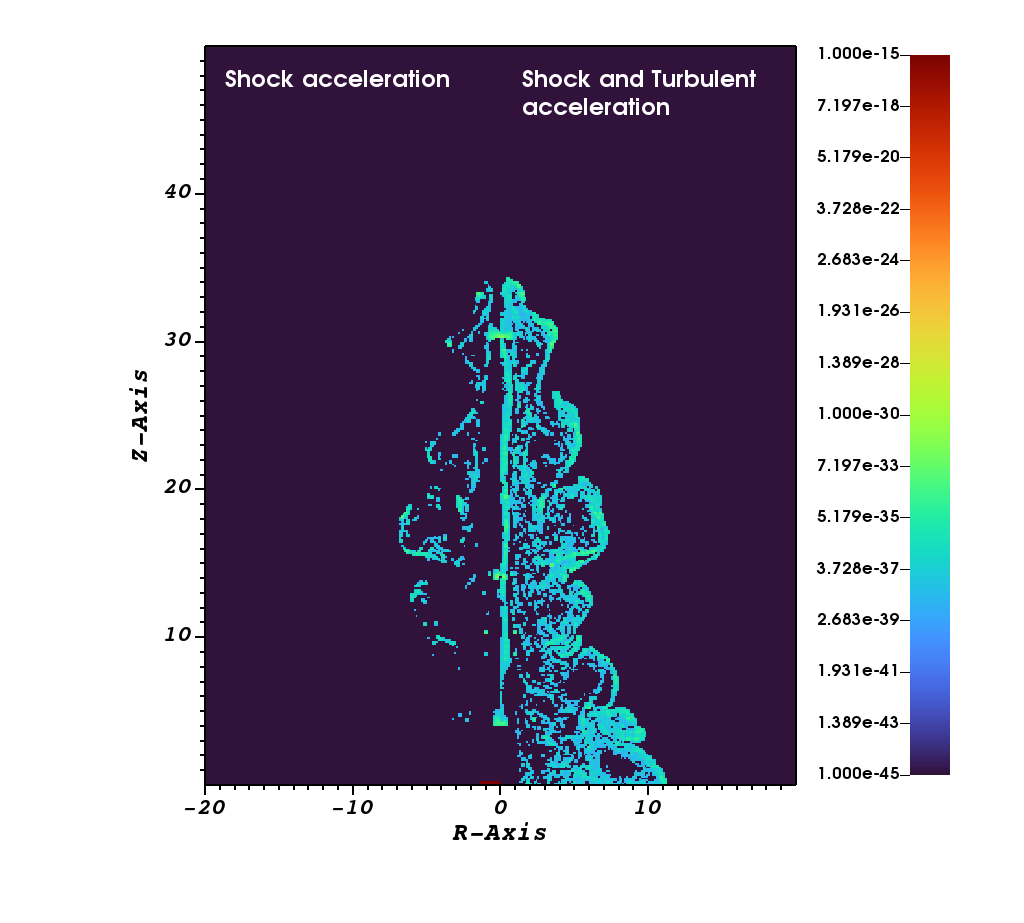}
    \vskip-2ex
    \caption{Same as Fig. \ref{fig:rad_1_4} but for $0.4\,KeV$ X-Ray at time $\tau=200$.}
    \label{fig:x_ray}
\end{figure}

In this section, we describe a toy model of a relativistic magneto-hydrodynamic jet and analyze its emission signatures due to the DSA and STA of cosmic rays. 
In particular, we employ a $2D$ cylindrical grid $\{R,Z\} \in \{0,0\}$ to $\{20,50\}$ using $160\times 400$ grid cells.
The ambient medium is initially static ($\vec{V}_{m}=0$) with constant density $\rho_{m}=10^{3} \hat{\rho}_{0}$, where, $\hat{\rho}_{0} = 1.67\times10^{-24}\,{\rm gr}\,{\rm cm}^{-3}$. 
An under-dense beam with $\rho_{j}=\hat{\rho}_{0}$ is injected into the ambient medium with velocity $v_{z}$  along the vertical direction through a circular nozzle of unit radius, $R_{j} = \hat{L}_{0}$ from the lower $Z$ boundary. 
The value of $v_{z}$ is prescribed using the Lorentz factor $\gamma_j=10$ and $\hat{L}_0 = 100\,{\rm pc}$ implying an unit timescale of $\hat{\tau}_{0}=326.4\,{\rm yrs}$. 
The magnetic field is purely poloidal, $\vec{B}= B_z\hat{\vec{e}}_z$ and is initially prescribed in jet nozzle and also in the ambient medium,

\begin{equation}
    B_{z}=\sqrt{2\sigma_{z}P_{j}}.
\end{equation}

where, $P_{j}$ is the jet pressure at $R=R_{j}$ estimated from the Mach number  $M=v_{j}\sqrt{\rho_{ j}/(\Gamma P_{ j})+1/(\Gamma - 1)}=6$ and adiabatic index $\Gamma=5/3$. 
The values for $\sigma_{z}$ is taken to be $10^{-4}$ for the present simulation. 

We further inject $25$ Lagrangian macro-particles every two time steps with an initial power-law spectral distribution with index $-9$ on a initial $\gamma$ grid with $\{\gamma_{\min},\gamma_{\max}\}\equiv\{1,10^{ 5}\}$ discretized with $128$ bins. 

The energy spectrum of the macro-particles are calculated for two different scenarios: i) considering only DSA and different losses and ii) considering, in addition, also stochastic processes.
For scenario (i) we follow the numerical algorithm developed in \cite{vaidya_2018, mukherjee_2021} to estimate the particle spectral distribution, while for scenario (ii) we solve Eq.~(\ref{eq:main}) without the source and sink terms, along with the diffusion coefficient $D\propto\gamma^2$, where the proportionality constant is computed from the value of $t_{ A}$ following Eq.~(\ref{eqn:main}) and with the value of $\lambda_{\max}=\hat{L}_{0}/100$. 
The loss terms account for synchrotron, Inverse Compton and adiabatic losses.
Also, compared to the previous test problems here we take Courant number $0.8$ when solving Eq.~(\ref{eq:main}). 
Moreover, for both scenarios we compute the emissivity for each macro-particle based on their local spectral distribution and interpolated it on the underlying grid \citep{vaidya_2018}.

In Fig.~\ref{fig:jet_spec}, we show the spectral evolution of representative particles, that have been shocked at least once, for each of the scenarios.
The top panel shows spectral evolution of a representative particle for the case where acceleration is due to shocks alone.
The effect of DSA and radiative losses are clearly visible, respectively, from the spectral flattening and from high energy cut-offs.
Here the cut-off can be observed clearly, as during DSA, the maximum energy get shifted according to the prescription described in Sec.~\ref{sec:turb_theory}. 
When the maximum $\gamma$ exceeds its initial value, cooling processes become effective so that the macro-particle  quickly cools accounting for sharp spectral cut-off.

The bottom panel shows the spectral evolution of similar particle for the case where STA is also included (besides DSA). 
the distribution reveals a hump-like structure in the low-energy end of the spectrum that slowly shifts towards higher $\gamma$ values.
With time, this eventually leads the distribution function to reach a steady state, as described by  Eq.~(\ref{eq:steady_q_2}).
Notice that our choice of parameters (Eq.~\ref{eqn:main}) is such that the acceleration timescale $t_A$ is larger or comparable to the dynamical time, leading to feeble acceleration.
We also point out that, during the initial stages, the particle spectrum exhibits a pile-up effect at low $\gamma$, because of the finite grid constraint, as discussed in section \ref{sec:hard_sp}.
This spurious effect dims with time as lower $\gamma$ particles starts to accelerate toward higher $\gamma$.
The impact of DSA (in addition to STA) can be distinguished from the flattening of the spectral distribution.
The more pronounced low-energy cutoff is attributed to the lower energy particles being accelerated by STA, eventually creating a deficiency in the number of particles at low $\gamma$.

From the instantaneous spectral distribution of Lagrangian macro-particles spread across the computational domain, we estimate the synchrotron emissivity by convolving the macro-particle spectra with single electron synchrotron spectra and interpolated it on the computational grid \citep[see Eq. 36-37 in][]{vaidya_2018}. 
In Figs.~\ref{fig:rad_1_4}, \ref{fig:optical} and \ref{fig:x_ray}, the emissivity $J_{\nu}$ computed from the Lagrangian macro-particles is shown for different frequencies at time $\tau=200\hat{\tau}_{0}$ for the two different scenarios (left and right halves, respectively).

In Fig.~\ref{fig:rad_1_4}, with $1.4\,{\rm GHz}$ radio frequency, the emission due to turbulence and shock (right half) is very similar to the case with DSA only (left half).
For the case with optical frequency ($\nu = 6.59\times 10^{ 5}$\, GHz) (Fig.~\ref{fig:optical}), the emission becomes less than the radio frequency (Fig.~\ref{fig:rad_1_4}) for both cases with and without STA. 
This is expected because of the faster cooling time with higher energy. 
However, a significant larger emission can be seen in case ii) in the region $Z \lesssim 10$.
The material in this region originates from the back-flow dynamics of the jet \citep{cielo_2014,matthews_2019}. 
If only shock energization is accounted for, the particle spectra become very steep in this region owing to radiative losses and the absence of strong shocks. 
However, if STA is also taken into account, the spectra remain hard because of the competing effects of STA and radiative losses. Similar high emission features are observed in X-ray ($\nu = 10^{ 8}$ GHz) as well (right panel of Fig.~\ref{fig:x_ray}).
On the contrary, in the presence of DSA only, a significant reduction in the X-ray emission can be seen (left half).
Here most of the emission originates from the regions near jet head as well as isolated spots in the cocoon.
In addition, smaller emission centers can be observed in the region around the re-collimation shocks along the beam.
This differs from the case with DSA + STA, where the emission pattern was wider and more uniformly distributed throughout the jet and the backflow region. 
%
\section{Discussion and Summary} \label{sec:dis}
%
%
In this paper we have focused on the numerical modeling of stochastic turbulent acceleration (STA) and its physical contribution to the spectral evolution of highly energetic particles.
The numerical formulation is based on the fluid-particle hybrid framework of \cite{vaidya_2018, mukherjee_2021} developed for the PLUTO code, where the non-thermal plasma component is modeled by means of Lagrangian macro-particles embedded in a classical or relativistic magnetized thermal flow.

The particle distribution function is evolved by solving numerically a Fokker-Planck equation in which STA is modelled by two components: a hyperbolic term describing the systematic acceleration (Fermi $\rm II$) and a parabolic contribution accounting for random resonant interaction between particles and plasma turbulent waves.
While \cite{vaidya_2018} presented a Lagrangian method for the solution of the Fokker-Planck equation in the presence of hyperbolic terms only, here we have introduced a novel Eulerian algorithm to account also for an energy-dependent diffusion coefficient $D\sim\gamma^2$ which can become stiff in the high-energy limit.
To overcome the explicit time step restriction, the new method takes advantage of $2^{\rm nd}$-order Runge Kutta Implicit-Explicit (IMEX) methods, so that hyperbolic terms (e.g. adiabatic expansion / radiative losses / Fermi $\rm II$) are treated explicitly while parabolic terms (modelling turbulent diffusion) are handled implicitly. 

Selected numerical benchmarks validated against analytical solutions and grid resolution studies demonstrate that our implementation has improved stability and accuracy properties when compared to previous solvers \citep[see for example][]{chang_1970,winner_2019}.
In addition, due to the presence of boundary condition our algorithm respects physical constraints (for example, $\gamma\geq1$) which are not always satisfied in the Lagrangian method \citep{vaidya_2018, mukherjee_2021} with an evolving grid.
STA modeling has also been validated against radiative synchrotron loss process by studying the evolution of curvature of particle spectrum \citep{tramacere_2011}.

With these motivations, we have studied the effect of STA as well as other energization processes, on the particle spectrum in the presence of shocks, using toy-model applications.
Such an interplay is commonly believed to operate in supernova remnants, AGN radio lobes, galaxy clusters and radio relics.

As a first application example, we considered a simple planar shock in four different acceleration scenarios. 
We found that when STA and DSA both are considered, the former seems to affect the shock injection by changing the  macro-particle distribution function.  
Further tests with different forms of the diffusion coefficient reveal a similar behavior. 
Additionally, we have also quantified the effect of STA time scale on the radiative losses and its influence on the interplay with DSA. In particular, we observe that the effect of shocks on particle distribution weakens with decreasing STA time scales. 
Similar interplay of DSA and STA was also evident in case of spherical shock formed in the test case of RMHD blast wave. 

Finally, we have extended our algorithm to explore the emission properties of the axisymmetric RMHD jet using a toy model. 
We find a significant difference both in the evolution of the spectral distribution and the ensuing emission signatures due to the presence or absence of the STA process.
In particular, inclusion of STA results in diffuse emission within the jet back-flow, particularly in the high-energy X-ray band. 
Consequences of such an important finding will be further explored in forthcoming works focusing on astrophysical systems along with comparison with observed signatures.

\section*{ACKNOWLEDGEMENTS}
We would like to thank the anonymous referee for the helpful comments, and constructive remarks on this manuscript. 
All simulations were performed at the computing facility at Indian Institute of Technology, Indore. 
We would like to thank the financial support from the Max Planck partner group award at Indian Institute of Technology, Indore.

\bibliographystyle{aasjournal}
\bibliography{ref_kundu}

\appendix
\section{Analytical solution of Fokker-Planck Equation} \label{sec:analytic}
%
%
Eq.~(\ref{eq:main}) is very hard to solve for a proper general analytic solution. Various work has been devoted to solve Eq.~(\ref{eq:main}) for various transport coefficients \citep[e.g.,][]{katarzynski_2006,park_1995,chang_1970,kardashev_1962}. \cite{chang_1970} solved Eq.~(\ref{eq:main}) for the steady-state solution and the solution could be written as,

\begin{eqnarray}
\chi_{\rm steady}(\gamma)=\chi_{\rm 0}\exp\Big\{-\int^{\gamma}_{\rm 1}\Big(\frac{S(\gamma',\tau)-D_{\rm A}(\gamma',\tau)}{D_{\gamma \gamma}(\gamma',\tau)}\Big)d\gamma'\Big\}.
\label{eq:chang}
\end{eqnarray}

\cite{katarzynski_2006} solved Eq.~(\ref{eq:chang}) for $D_{\gamma \gamma}(\gamma,\tau)=D_{\gamma \rm 0}\gamma^{\rm 2}/2$ with $D_{\gamma \rm 0}=1/t_{\rm A}$, $D_{\rm A}(\gamma,\tau)=\gamma/t_{\rm A}$ and $S(\gamma,\tau)=S_{\rm 0}\gamma^{\rm 2}$.
These form of the parameters are typical for particles in plasma. The loss term $S(\gamma,\tau)$ gets a similar form if Inverse-Compton radiation is taken in the Thompson limit with Synchrotron radiation and the form for the diffusion coefficient $D_{\gamma \gamma}$ which also matches the form from typical particle in cell simulation as discussed above. The solution to Eq.~(\ref{eq:chang}) with the above mentioned parameters is,

\begin{eqnarray}\label{eq:steady_q_2}
\chi_{\rm steady}(\gamma)=\chi_{\rm 0}\gamma^{\rm 2}\exp\{-2S_{\rm 0}t_{\rm A}(\gamma-1)\}.
\end{eqnarray}

\cite{kardashev_1962} got a time-dependent solution for Eq.~(\ref{eq:main}) without the loss terms and showed the acceleration leads to a log-normal particle distribution (similar to Eq.~(\ref{eq:fit})). 

So, if the particles only accelerate via STA the particle distribution follows a log-normal form due to the fact that the STA process is a multiplicative acceleration process \citep{tramacere_2011}. But if those particles loose their energy via radiative means along with the acceleration the particle distribution starts to follow an ultra-relativistic Maxwellian (Eq.~(\ref{eq:steady_q_2})), which looks like a thermal or quasi-thermal spectrum with a scaled temperature of $1/S_{\rm 0}t_{\rm A}$ which is also the value of $\gamma$ where, $t_{\rm A}=t_{\rm L}$.
\end{document}